\documentclass[%
 reprint,
superscriptaddress,
 amsmath,amssymb,
 aps,
onecolumn,
showkeys,
]{revtex4-2}
\usepackage{graphicx}
\usepackage{dcolumn}
\usepackage{bm}

\begin{document}

\title{Charmonium: Conventional and $XYZ$ States in a \\ Relativistic Screened Potential Model}

\author{Chaitanya Anil Bokade}
 \email{anshul497@gmail.com}
\author{Bhaghyesh}%
\email{bhaghyesh.mit@manipal.edu; Corresponding author}
 \altaffiliation[ORCID iD: ]{0000-0003-3994-9945}
\affiliation{Department of Physics, Manipal Institute of Technology\\Manipal Academy of Higher Education, Manipal, 576104, Karnataka, India}

\begin{abstract}
In this work a comprehensive analysis of the mass spectrum and decay properties of charmonium states within a relativistic framework is carried out. The present experimental status of charmonium states is reviewed. Utilizing a screened potential model, we compute the spectra and various decay widths for $c\bar{c}$ bound system, comparing our results with experimental data and existing theoretical models. We calculate the decay constants, $E1$ and $M1$ transitions, and annihilation decay widths, confirming the consistency of our model with experimental observations for well-established charmonium states. The interpretation of charmonium-like states, $X(4140)$, $X(4274)$, and $X(4500)$ as $P$ wave charmonium states and $\psi(4040)$, $\psi(3770)$, $\psi(4160)$, $\psi(4230)$, $\psi(4360)$, $\psi(4415)$, $Y(4500)$ and $Y(4360)$ as $S-D$ mixed charmonium states are carried out.


\end{abstract}
\maketitle


\section{\label{sec:Introduction} Introduction}

Heavy-flavour hadron spectroscopy has always been of great interest to theoretical and experimental physicist as it allows to test fundamental properties of QCD. This interest has grown ever since the discovery of $J/\psi$ state of charmonium \cite{1}. All the charmonium resonance below the $D^{0}\bar{D}^{0}$ are experimentally measured with high accuracy \cite{exp} and are also well described theoretically with the Cornell potential \cite{3,4}. However, for higher states, there are disagreement with experimental results \cite{4}. Even with these discrepancies some of these higher resonances ($\psi(4040)$, $\psi(4160)$ and, $\psi(4415)$) are still interpreted in terms of conventional charmonium states. Following the first discovery of $X(3872)$ at Belle in 2003 \cite{5}, an array of charmonium-like states were identified. These exotic states contains $c\bar{c}$ structure but showed properties different compared to $c\bar{c}$ meson. In 2013, Belle collaboration from the decay processes $B\rightarrow K\gamma\chi_{c1}$ and $B\rightarrow K\gamma\chi_{c2}$  observed a narrow signal of a new resonance $\psi(3823)$ \cite{6} and suggested it to be  $\psi_{2}(1D)$ charmonium state. Later  in 2015, BESIII collaboration observed clear peak for  $\psi(3823)$ with mass of $3821.7\pm1.3\pm0.7$ MeV in the  $e^{+}e^{-}\rightarrow \pi^{+} \pi^{-} \psi(3823)\rightarrow \pi^{+} \pi^{-} \gamma\chi_{c1}$ decay and confirmed it to be $\psi_{2}(1D)$ \cite{7}. Recently in 2019, LHCb collaborations observed a resonance $X(3842)$ for first time in $X(3842)\rightarrow D^{0}\bar{D}^{0}$ and $X(3842)\rightarrow D^{+}D^{-}$ decay modes and predicted it to be the unobserved $\psi_{3}(1D)$ charmonium state \cite{8}. In the examination of $\gamma\gamma\rightarrow \omega J/\psi$ decay process, Belle collaboration observed $X(3915)$ \cite{9}, which was later confirmed by BaBar in the same decay process with mass $3919.4\pm2.2\pm1.6$ MeV, decay width of $13\pm6\pm3$ MeV and spin-parity of $J^{P}= 0^{+}$ , suggesting $X(3915)$ to be $\chi_{c0} (2P)$ charmonium state \cite{10}. PDG formerly considered $X(3915)$ to be $\chi_{c0} (2P)$, but this assignment was later taken down as it faced some challenges, which has been discussed in ref. \cite{11,12,13}. Later in 2017, when Belle collaboration observed $X(3860)$ for first time while analyzing the full wave amplitude of $e^{+}e^{-}\rightarrow J/\psi D\bar{D}$ decay process, having mass of $3862_{-32-13}^{+26+40}$ MeV, width $201_{-67-82}^{+154+88}$ MeV and $J^{PC}= 0^{++}$ \cite{14}. $X(3860)$ was considered to be more suited for $\chi_{c0}(2P)$ charmonium state \cite{14}. $X(3930)$ was first observed by Belle in 2006 \cite{15} and later by BaBar in 2010 \cite{16} in $\gamma\gamma\rightarrow D\bar{D}$ decay process and both suggested  $X(3930)$ to be  $\chi_{c2}(2P)$ charmonium state. In the analysis of $e^{+}e^{-}\rightarrow J/\psi$ decay mode Belle collaboration discovered the $X(3940)$ resonance \cite{17}. They later measured its mass and width to be $3942_{-6}^{+7}\pm6$ MeV and $37_{-18}^{+26}\pm8$ MeV, respectively in the decay process $e^{+}e^{-}\rightarrow J/\psi D^{*}\bar{D}^{*}$ \cite{18}. They also observed $X(4160)$ with the mass and decay width of $4156_{-20}^{+25}\pm15$ MeV and $139_{-61}^{+111}\pm21$ MeV ,respectively \cite{18}. There is no clear understanding about the nature of $X(3940)$ and $X(4160)$ because of limited experimental data but the most common interpretation in the literature for $X(3940)$ is the $\eta_{c}(3S)$ charmonium state \cite{19,20,21}, while the $X(4160)$ is suggested to be $\eta_{c}(4S)$ or $\chi_{c0}(3P)$ in \cite{22}. Although, the $\eta_{c}(4S)$ interpretation of $X(4160)$ is not favored in ref. \cite{20,21}. $X(4140)$ was first reported by CDF experiments in 2009 as a narrow structure near $J/\psi\phi$ threshold by analyzing the data of $B^{+}\rightarrow J/\psi\phi K^{+}$ decays \cite{23}. In 2010, Belle conducted an analysis of $\gamma\gamma\rightarrow J/\psi\phi$ process in search of $X(4140)$, but instead they observed $X(4350)$ with mass of $4350.6_{-5.1}^{+4.6}\pm0.7$ MeV and decay width $13_{-9}^{+18}\pm4$ MeV, respectively, with $0^{++}$ or $2^{++}$ as the possible assignment of $J^{PC}$ \cite{24}. When CDF collaboration investigated $B^{+}\rightarrow J/\psi\phi K^{+}$ decay process to confirm $X(4140)$ using a larger data sample, they discovered evidence of $X(4274)$ with mass and width of $4274.4_{-6.7}^{+8.4}\pm1.9$ MeV and $32.3_{-15.3}^+{21.9}\pm7.6$ MeV, respectively \cite{25}. In 2017, LHCb collaboration analyzed the data for $B^{+}\rightarrow J/\psi\phi K^{+}$  events and observed 4 possible resonant structures, two were the previously observed $X(4140)$ and $X(4274)$ to which they assigned the  $J^{PC}= 1^{++}$ \cite{26}. Along with two new resonances, $X(4500)$ with mass and width of $4506\pm11_{-15}^{+12}$ MeV and $92\pm 21_{-20}^{+21}$ MeV, respectively, and $X(4700)$ with mass and width of $4704\pm10_{-24}^{+14}$ MeV and $120\pm31_{-33}^{+42}$ MeV, respectively both having  $J^{PC}= 0^{++}$ \cite{26}. In 2021, LHCb again analyzed the $B^{+}\rightarrow J/\psi\phi K^{+}$ decay process with six times the sample data and confirmed the previously observed $X(4140)$, $X(4274)$, $X(4500)$ and $X(4700)$ states \cite{27}. Additionally they also observed two more structures, $X(4685)$ with mass and width of $4684\pm7_{-16}^{+13}$ MeV and $126\pm15_{-41}^{+37}$ MeV, respectively having  $J^{P}= 1^{+}$ and $X(4630)$ with mass and width of $4626\pm16_{-110}^{+18}$ MeV and $174\pm27_{-73}^{+134}$ MeV, respectively with preferred  $J^{P}= 1^{-}$ \cite{27}. $Y(4260)$ was first observed by BaBar in their study of $e^{+}e^{-}\rightarrow \pi^{+} \pi^{-}J/\psi$ via ISR process as a broad structure at $4.26$ GeV \cite{28} which was later confirmed by CLEO collaboration \cite{29} and Belle \cite{30} in the same decay channel. In the same process Belle were first to observe another broad structure named $Y(4008)$ \cite{30} which was again confirmed by them in 2013 with larger sample data \cite{31}. BESIII collaboration recently conducted precise measurement of $e^{+}e^{-}\rightarrow \pi^{+} \pi^{-} J/\psi$ process and confirmed $Y(4260)$ with mass and decay width of $4222.0\pm3.1\pm1.4$ MeV and $44.1\pm4.3\pm2.0$ MeV, respectively \cite{32}. They also discovered another structure called $Y(4320)$ with mass and decay width of $4320.0\pm10.4\pm7.0$ MeV and $101.4_{-19.7}^{+25.3}\pm10.2$ MeV respectively, but no observation of resonant structure $Y(4008)$ was reported \cite{32}. BESIII collaboration searched for the production of $e^{+}e^{-}\rightarrow \omega\chi_{cJ}$, $(J = 0,1,2)$ decay modes and analyzed $\omega\chi_{c0}$ cross section in CM frame detecting a resonance $Y(4230)$, inconsistent with previously observed $Y(4260)$ \cite{33}. The measured $Y(4230)$ resonance had mass and width of $4230\pm8\pm6$ MeV and $38\pm12\pm2$ MeV, respectively \cite{33}. BESIII collaboration again observed $Y(4230)$ in cross sections of $e^{+}e^{-}\rightarrow K^{+} K^{-} J/\psi$ along with a new structure $Y(4500)$ with significance greater than $8\sigma$, having mass and width of $4484\pm13.3\pm24.1$ MeV and $111.1\pm30.1\pm15.2$ MeV, respectively \cite{34}. In BESIII collaboration analysis of $e^{+}e^{-}\rightarrow \pi^{+} \pi^{-}h_{c}$  process cross section, they observed two resonance structures, the $Y(4220)$, which was later confirmed by \cite{35}, and the $Y(4390)$ \cite{36}. BaBar \cite{37} and Belle \cite{38} both studied the $e^{+}e^{-}\rightarrow \pi^{+} \pi^{-}\psi(2S)$ process via ISR and observed two resonant structures $Y(4360)$ and $Y(4660)$ in the invariant mass distribution of $\pi^{+} \pi^{-}\psi(2S)$. Belle studied the  $e^{+}e^{-}\rightarrow \Lambda_{c}^{+}\Lambda_{c}^{-}$ in the ISR process and observed a resonant structure called $Y(4630)$ for the first time, having mass and decay width of $4634_{-7-8}^{+8+5}$ MeV and $92_{-24-21}^{+40+10}$ MeV, respectively \cite{39}. In 2008, Belle reported the first charmonium-like charged structure $Z_{c}(4430)$ \cite{40}. $Z_{c}(3900)$ state was observed by BESIII \cite{41} and Belle \cite{31} in 2013. BESIII searched for $Z_{c}(3900)$ in $e^{+}e^{-}\rightarrow \pi^{+} \pi^{-}h_{c}$ process but observed $Z_{c}(4020)$ instead \cite{42}. These charmonium-like charged particles cannot be described as the conventional $q\overline{q}$ meson structures. Collectively these exotic particles which are charmonium-like but show properties different to conventional charmonium states are called as $XYZ$ particles. To understand the underlying structure of these $XYZ$ particles there are many interpretations in the literature, such as hadronic molecules, hybrids, hadro-quarkonia, glueballs, tetraquarks, pentaquarks, etc \cite{43,44,45,46}. It is still unclear how to understand the entire spectrum of charmonium-like structures. For better understanding and interpretation of the unconventional nature of $XYZ$ states, the knowledge of charmonium serves as a foundation due to their simpler internal structures and precise measurements below the open charm threshold, which makes them better understood compared to $XYZ$ states. It is imperative to study them simultaneously as both, conventional and exotic states complement each other \cite{47}. Given the experimental progress and discoveries in charmonium spectroscopy, the need for improved theoretical models becomes evident.

In this article we have conducted a comprehensive study of charmonium in a relativistic screened potential model. In section \ref{sec:Methodology}, we discuss the theoretical model used to describe the charmonium bound system and the numerical approach used to solve the relativistic Schrodinger equation. Decay constants are discussed in section \ref{sec:Decay constant}. In sections \ref{sec:Radiative Transitions} and \ref{sec:Annihilation Decays} presents the various decays that are investigated. In section \ref{sec:S-D Mixing} $S-D$ mixing of charmonium states is discussed. In section \ref{sec:Results and Discussion} a thorough investigation of our evaluation and interpretation of $XYZ$ states are conducted, along with comparison between the experimental results and theoretical models. In section \ref{sec:Conclusion} we present our conclusion.

\section{\label{sec:Methodology} Methodology}

To compute the mass spectra of charmonium with relativistic kinematics we have used the relativistic generalization of the Schrodinger equation also called the spinless Salpeter equation \cite{4}

\begin{equation}
	\label{eq:1}
	H = \sqrt{-\nabla_{q}^{2} + m_{q}^{2}}+\sqrt{-\nabla_{\bar{q}}^{2} + m_{\bar{q}}^{2}} + V(r) ,
\end{equation}
where $\vec{r}= \vec{x}_{\bar{q}}-\vec{x}_{q}$, where $\vec{x}_{\bar{q}}$ and $\vec{x}_{q}$  are the coordinates of the quarks and operators $\nabla_{q}^{2}$ and  $\nabla_{\bar{q}}^{2}$ are the partial derivative of those coordinates respectively. $m_{q}$  and $m_{\bar{q}}$ are the masses of quark and an anti-quark respectively. The interaction potential $V(r)$ between the quark and antiquark is described by a screened potential \cite{48}
\begin{equation}
	V(r) = V_{V}(r)+V_{S}(r),
\end{equation}
\noindent where,
\begin{equation}
	V_{V}(r) = -\frac{4}{3}\frac{\alpha_{s}(r)}{r} \,,
	\qquad
	V_{S}(r) = \lambda\left(\frac{1 - e^{-\mu r}}{\mu}\right) + V_{0} \label{potterms}.
\end{equation}

\noindent $V_{V}(r)$ is the one-gluon-exchange coulomb potential term which is dominant at short distance. The second term $V_{S}(r)$ is the confining term modified to consider colour screening effect which is dominant at long distances. A screened potential can account for the effects of virtual quark-antiquark pairs that become significant at higher energies, providing a more accurate description of the quark-antiquark interactions and can resolve the discrepancies observed in higher charmonium states \cite{48}. $\mu$ is called as the screening factor which makes the long-range part of $V(r)$ flat when $r \gg \frac{1}{\mu}$, and linearly rising when $r \ll \frac{1}{\mu}$, $\lambda$ is the linear potential slope. As $\mu \rightarrow 0$, $V(r)$ reduces to Cornell potential \cite{48}. In Eq. \ref{potterms}, $\alpha_{s}(r)$ is the running coupling constant. The running coupling constant in coordinate space can be obtained by Fourier transformation of coupling constant in momentum space $\alpha_{s}\left(Q^{2}\right)$ \cite{4} and is given by,
\begin{equation}
	\alpha_{s}(r) = \sum_{i}\alpha_{i}\frac{2}{\sqrt{\pi}}\int_{0}^{\gamma_{i}r}e^{-x^{2}}dx ,
\end{equation}
where $\alpha_{i}'s$ are the free parameters fitted to replicate the short distance behavior of $\alpha_{s}\left(Q^{2}\right)$ predicted by QCD. The numerical values for the parameters are $\alpha_{1}=0.15$, $\alpha_{2}=0.15$, $\alpha_{3}=0.20$, and $\gamma_{1}=1/2$, $\gamma_{2}=\sqrt{10}/2$, $\gamma_{3}=\sqrt{1000}/2$ \cite{4,49}. The spin dependent interaction potential which removes the degeneracy between the states is given by
\begin{equation}
	V_{SD}(r) = V_{SS}(r)\vec{S}_{q}\cdot \vec{S}_{\bar{q}}+V_{LS}(r)\vec{L}\cdot \vec{S}+V_{T}(r)S_{12} ,
\end{equation}
where the spin-spin term $V_{SS}$ gives the spin singlet-triplet hyperfine splitting.  The spin orbit term $V_{LS}$ and the tensor term $V_{T}$, together describe the fine structure of the states. $V_{SS}$, $V_{LS}$, and $V_{T}$ are represented by \cite{50,51}
\begin{align}
V_{SS}(r) &= \frac{32\pi\alpha_{s}(r)}{9m_{q}^{2}}\tilde{\delta}_{\sigma}(r)\,,
\nonumber \\
V_{LS}(r) &= \frac{1}{2m_{q}^{2}r}\left(3V_{V}^{'}(r)-V_{S}^{'}(r)\right) \,,
\nonumber \\
V_{T}(r) &= \frac{1}{2m_{q}^{2}}\left(\frac{V_{V}^{'}(r)}{r}-V_{V}^{''}(r)\right) \,. 
\end{align}
Here, $\tilde{\delta}_{\sigma}(r) = \left(\frac{\sigma}{\sqrt{\pi}}\right)^{3}e^{-\sigma^{2}r^{2}}$ is the smeared delta function \cite{52,53,54}. These spin-dependent interactions are diagonal in the $|J,L,S\rangle$ basis with matrix elements given by \cite{53,54,55}
\begin{equation}
	\begin{split}
		\langle\vec{S}_{q}\cdot \vec{S}_{\bar{q}}\rangle &= \frac{1}{2}S^{2}-\frac{3}{4} \,,
		\\
		\langle\vec{L}\cdot \vec{S}\rangle &= \left(J(J+1)-L(L+1)-S(S+1)\right)/2 \,. 
	\end{split}
\end{equation}
The tensor operator $S_{12}=3\left(\vec{S}_{q}\cdot\hat{r} \right)\left(\vec{S}_{\bar{q}}\cdot\hat{r}\right)- \vec{S}_{q}\cdot\vec{S}_{\bar{q}}$, has non-vanishing diagonal matrix elements only between $L > 0$ spin-triplet states:
\begin{equation}
	\langle S_{12} \rangle= \begin{cases}
		-\frac{L}{6(2L+3)} & J=L+1 
		\\
		\frac{1}{6} & J=L
		\\
		-\frac{L+1}{6(2L-1)} & J=L-1 \,.
	\end{cases} 
\end{equation}

To obtain the spin-averaged masses, the Schrodinger equation is solved as an eigenvalue equation using the method developed in Refs. \cite{49,56}. A brief summary of the method is given below. The relativistic wave equation corresponding to the Hamiltonian \eqref{eq:1} is given by
\begin{equation}
	\left(\sqrt{-\nabla_{q}^{2} + m_{q}}+\sqrt{-\nabla_{\bar{q}}^{2} + m_{\bar{q}}} + V(r)\right)\Psi(\vec{r})  = E\Psi(\vec{r}) \label{waveeq}.
\end{equation}
Expanding the wave function in terms of spectral integration,
\begin{equation}
	\Psi(\vec{r}) = \int d^{3}r'\int \frac{d^{3}k}{(2\pi)^{3}}e^{i\vec{k}(\vec{r}-\vec{r}')}\Psi(\vec{r}') ,
\end{equation}
we can rewrite \eqref{waveeq} as
\begin{widetext}
\begin{equation}
	\label{eq:2}
	V(r)\Psi\left(\vec{r}\right)+\int d^{3}r'\frac{d^{3}k}{(2\pi)^{3}}\left(\sqrt{k^{2} + m_{q}}+\sqrt{k^{2} + m_{\bar{q}}}\right)e^{i\vec{k}(\vec{r}-\vec{r}')}\Psi\left(\vec{r}'\right) = E\Psi\left(\vec{r}\right) .
\end{equation}
\end{widetext}
The exponential term in the above equation can be decomposed in terms of spherical harmonics as
\begin{equation}
	e^{i\vec{k}\cdot\vec{r}} = 4\pi\sum_{nl}Y_{nl}^{*}\left(\hat{k}\right)Y_{nl}(\hat{r})j_{l}(kr)i^{l} \label{expterm},
\end{equation}
where $j_{l}$ is the spherical Bessel function, $Y_{nl}^{*}(\hat{k})$ and $Y_{nl}(\hat{r})$ are the spherical harmonics with normalization condition $\int d\Omega Y_{n_{1}l_{1}}(\hat{k})Y_{n_{2}l_{2}}(\hat{r}) = \delta_{n_{1}n_{2}}\delta_{l_{1}l_{2}}$, $\hat{k}$ and $\hat{r}$ are unit vectors along the $\vec{k}$ and $\vec{r}$ direction, respectively. Factoring the wave function into radial $R_{l}(r)$ and angular $Y_{nl}(r)$ parts, and substituting \eqref{expterm} in \eqref{eq:2} and simplifying, we get \cite{49,56}

\begin{widetext}
	\begin{equation}
		\label{eq:3}
		V(r)u_{l}(r) + \frac{2}{\pi}\int dkk^{2}\int dr'rr'\left(\sqrt{k^{2} + m_{q}^{2}}+\sqrt{k^{2} + m_{\bar{q}}^{2}}\right)j_{l}(kr)j_{l}(kr')u_{l}(r') = Eu_{l}(r),
	\end{equation}
\end{widetext}

where $u_{l}(r)$ is the reduced radial wave function ($R_{l}(r) =  u_{l}(r)/r$).

\noindent For a quark-antiquark bound state, as the separation distance increases, the wavefunction significantly diminishes. At sufficiently large distances, the wavefunction effectively drops to zero. To characterize this behaviour, a typical distance scale, denoted as $L$, is introduced. This effectively confines the bound state's wavefunction within the spatial range of $0<r<L$. Then the reduced wavefunction $u_l(r)$ can be expanded in terms of spherical Bessel function for angular momentum $l$ as

\begin{equation}
	\label{eq:4}
	u_l(r) = \sum_{n=1}^{\infty}c_{n}\frac{a_{n}r}{L}j_{l}\left(\frac{a_{n}r}{L}\right) ,
\end{equation}
where $c_{n}$'s are the expansion coefficients, $a_{n}$ are the $n$-th root of the spherical Bessel function, $j_{l}(a_{n})=0$. For large value of $N$, series \eqref{eq:4} can be truncated. The momentum is discretized as a result of confinement of space, which allows us to replace $a_{n}⁄L\leftrightarrow k$  and the integration in eq. \eqref{eq:3} can be replaced by $\int dk \rightarrow \sum_{n}\Delta a_{n}⁄L$, where,  $\Delta a_{n}= a_{n} - a_{n-1}$. For the limited space, $0<r,r'<L$, incorporating all the changes in the eq. \eqref{eq:3}, we get the final equation in terms of the coefficients $c_{n}$'s as \cite{49,56}

	\begin{eqnarray}
		\label{eqfinal}
		Ec_{m} &=&
		\sum_{n=1}^{N}\frac{a_{n}}{N_{m}^{2}a_{m}}\int_{0}^{L}drV(r)r^{2}j_{l}\left(\frac{a_{m}r}{L}\right)\left(\frac{a_{n}r}{L}\right)c_{n} \\ &+&\frac{2}{\pi L^{3}}\Delta a_{m}a_{m}^{2}N_{m}^{2}\left(\sqrt{\left(\frac{a_{m}}{L}\right)^{2} + m_{q}^{2}}+\sqrt{\left(\frac{a_{m}}{L}\right)^{2} + m_{\bar{q}}^{2}}\right)c_{m}
	\end{eqnarray}

where $N_{m}$ is module of spherical Bessel function:
\begin{equation}
	N_{m}^{2} = \int_{0}^{L}dr'r'^{2}j_{l}\left(\frac{a_{m}r'}{L}\right)^{2} \,.
\end{equation}
Eq. \eqref{eqfinal} is an eigenvalue equation in the matrix form, which is solved numerically. For large enough value of $L$ and $N$ the solution is nearly stationary \cite{49,56}.  The eigenvalues correspond masses of spin averaged states and the eigenvectors correspond to the wavefunction of these states. The spin dependent interactions are incorporated perturbatively using the normalized wavefunctions. The parameters used in our model are presented in Table \ref{tab:1} and are determined using the chi-square fitness test.
\begin{table}
	\caption{\label{tab:1} Parameters used in out model}
\begin{ruledtabular}
	\begin{tabular}{ccccc}
		$m_{q}$ (GeV) & $\sigma$ ($\text{GeV}^{2}$) & $\lambda$ (GeV) & $\mu$ (GeV) & $\Lambda$ (GeV)
		\\
		\colrule
		1.319 & 1.281 & 0.297 & 0.141 & 0.17\\
	\end{tabular}
\end{ruledtabular}
\end{table}
The masses of $S, P, D, F\ \text{and}\ G$ - wave states for $c\bar{c}$ bound system are presented in Tables \ref{tab:2}-\ref{tab:5}, respectively.

\section{\label{sec:Decay constant} Decay Constants}

Decay constants measure the amplitude of quark-antiquark annihilation into lighter particles. The decay constant of pseudoscalar $(f_{P})$ and vector $(f_{V})$ states can be calculated using the Van Royen Weisskopf formula \cite{57}
\begin{equation}
	f_{P/V} = \sqrt{\frac{3|R_{P/V}(0)|^{2}}{\pi M_{P/V}}}\bar{C}(\alpha_{s}),
\end{equation}
where $R_{P/V}(0)$ is the radial wavefunction at the origin for pseudoscalar (vector) meson state, $M_{P/V}$ is the mass of the pseudoscalar (vector) meson state and $\bar{C}(\alpha_{s})$ is the QCD correction given by \cite{58}
\begin{equation}
	\bar{C}^{2}(\alpha_{s}) = 1 - \frac{\alpha_{s}(\mu)}{\pi}\left(\delta^{P,V} - \frac{m_{q} - m_{\bar{q}}}{m_{q} + m_{\bar{q}}}\ln{\frac{m_{q}}{m_{\bar{q}}}}\right),
\end{equation}
where $\delta^{P}=2$ and $\delta^{V}=8/3$. The decay constant of $P-$wave states can be evaluated using \cite{59,60}

\begin{equation}
\begin{split}
	f_{\chi_{0}} &= 12\sqrt{\frac{3}{8\pi m_{q} }}\left(\frac{|R'_{\chi_{0}}(0)|}{M_{\chi_{0}}}\right) \,,
	\\
	f_{\chi_{1}} &= 8\sqrt{\frac{9}{8\pi m_{q} }}\left(\frac{|R'_{\chi_{1}}(0)|}{M_{\chi_{1}}}\right) \,.
\end{split}
\end{equation}
Here $M_{\chi_{0}}$ and $M_{\chi_{1}}$ are the masses of $\chi_{0}$ and $\chi_{1}$ states, respectively. The decay constants for pseudoscalar $f_{P}$ and vector $f_{V}$ are presented in Table \ref{tab:6} and decay constants for $f_{\chi_{0}}$ and $f_{\chi_{1}}$ are presented in Table \ref{tab:7}.

\section{\label{sec:Radiative Transitions} Radiative Transition}

Charmonium states characterized by higher quantum numbers exhibit a more compact nature owing to the substantial mass of their constituent quarks. Radiative transitions offer a distinctive avenue for probing the internal structure, since the initial $i$ and final $f$, $(i \rightarrow \gamma+f)$ states influence the radiative transitions. It offers a valuable means of detection, particularly for states with higher quantum numbers that prove challenging to observe through conventional methods. The discussion below centres on leading-order electromagnetic transitions, specifically focusing on the electric dipole $(E1)$ and magnetic dipole $(M1)$ transitions.

\subsection{\label{sec:E1 Transitions} E1 Transitions}

The $E1$ radiative partial widths between the states $\left(n_{i}^{2S+1}L_{J_{i}}^{i} \rightarrow \gamma+n_{f}^{2S+1}L_{J_{f}}^{f}\right)$ are given by \cite{61,62}
\begin{equation}
	\label{e1}
	\Gamma_{E1}\left(i \rightarrow \gamma+f\right) = \frac{4\alpha e_{q}^{2}}{3}E_{\gamma}^{3}
	\frac{E_{f}}{M_{i}}C_{fi}\delta_{S_{f}S_{i}}|\epsilon_{fi}|^{2}. 
\end{equation}
Here $\alpha=1/137$ is the fine structure constant,  $e_{q}$ is the quark charge, $E_{f}$ is the energy of the final state, $M_{i}$ is the mass of the initial state, $E_{\gamma}= \left(M_{i}^{2}-M_{f}^{2}\right)⁄2M_{i}$ is the emitted photon energy, $M_{f}$ is the mass of the final state, $E_{f}⁄M_{i}$  is the relativistic phase factor and $C_{fi}$ is the statistical factor given by 
\begin{equation}
	C_{fi} = \max\left(L_{i},L_{f}\right)\left(2J_{f}+1\right)
	\left\{
	\begin{array}{ccc}
		J_{i} & 1 & J_{f} \\
		L_{f} & S & L_{i} \\
	\end{array}
	\right\}^{2} ,
\end{equation}
where $\{:::\}$ are the $6j$ symbol. In \eqref{e1}, $\epsilon_{fi}$ is the overlapping integral determined using the initial $R_{n_{i}l_{i}}(r)$ and final state $R_{n_{f}l_{f}}(r)$ wavefunctions:
\begin{eqnarray}
	\epsilon_{fi} =&& \frac{3}{E_{\gamma}}\int_{0}^{\infty}drR_{n_{i}l_{i}}(r)R_{n_{f}l_{f}}(r)\nonumber\\
	&&\times
	\left(\frac{E_{\gamma}r}{2}j_{0}\left(\frac{E_{\gamma}r}{2}\right)-j_{1}\left(\frac{E_{\gamma}r}{2}\right)\right)
\end{eqnarray}
The $E1$ transitions widths for $S, P, D, F\ \text{and}\ G$ wave states are presented in Tables \ref{tab:8}-\ref{tab:11}, respectively.

\subsection{\label{sec:M1 Transitions} M1 Transitions}

The $M1$ radiative partial widths between the states $\left(n_{i}^{2S_{i}+1}L_{J_{i}} \rightarrow \gamma+n_{f}^{2S_{f}+1}L_{J_{f}}\right)$ are given by \cite{53,61,63}
\begin{equation}
	\Gamma_{M1}\left(i \rightarrow \gamma+f\right) = \frac{4\alpha \mu_{q}^{2}}{3}\frac{2J_{f}+1}{2L+1}E_{\gamma}^{3}
	\frac{E_{f}}{M_{i}}\delta_{L_{f}L_{i}}\delta_{S_{f}S_{i}\pm1}|m_{fi}|^{2},
\end{equation}
where $m_{fi}$ is the overlap integral 
\begin{equation}
	m_{fi} = \int_{0}^{\infty}drR_{n_{i}l_{i}}(r)R_{n_{f}l_{f}}(r)\left(j_{0}\left(\frac{E_{\gamma}r}{2}\right)\right),
\end{equation}
and $\mu_{q}$ is the magnetic dipole moment  given by  \cite{64}
\begin{equation}
	\mu_{q} = \frac{m_{\bar{q}}e_{q}-m_{q}e_{\bar{q}}}{2m_{q}m_{\bar{q}}} \,.
\end{equation}
The $M1$ transitions widths for $S$ and $P$ wave states are presented in Table \ref{tab:12}.

\section{\label{sec:Annihilation Decays} Annihilation Decays}

It is difficult to directly detect higher charmonium states because of its short life span, but it decays into extremely distinct final states that include photons, leptons, gluons, and quarks, which makes the analysis of these annihilation decays very crucial. The following section outlines and discusses various annihilation decays explored in this study.

\subsection{\label{sec:Di-Leptonic Decays} Di-Leptonic Decays}

Leptonic decay rates are essential for understanding the creation mechanism of charmonium resonance and can aid in differentiating between typical mesonic structures and more exotic or multi-quark structures. The charmonium resonance can decay through leptons $(l^{+}l^{-})$ via a virtual photon. The leptonic decay widths of $S$ - wave $(n^{3}S_{1})$  and $D$ - wave $(n^{3}D_{1})$  states are calculated using the Van Royen-Weisskopf formula (with the QCD correction factors) and are given by \cite{57,65,66,67}

\begin{equation}
	\begin{split}
		\Gamma\left(n^{3}S_{1} \rightarrow l^{+}l^{-}\right) &= \frac{4\alpha^{2}e_{q}^{2}}{M(n^{3}S_{1})^{2}}|R_{nS}(0)|^{2}\left(1-\frac{16\alpha_{s}(\mu)}{3\pi}\right) \,,
		\\
		\Gamma\left(n^{3}D_{1} \rightarrow l^{+}l^{-}\right) &= \frac{25\alpha^{2}e_{q}^{2}}{2m_{q}^{4}M(n^{3}D_{1})^{2}}|R''_{nD}(0)|^{2}\left(1-\frac{16\alpha_{s}(\mu)}{3\pi}\right) ,
	\end{split}
\end{equation}
where $R'_{nL}(0)$ is the value of radial wavefucntion at origin for $nL$ state and $(')$ represents the order of derivative and $M(n^{2S+1}L_{J})$ is the mass of $n^{2S+1}L_{J}$ state.

\subsection{\label{sec:Di-Photonic Decays} Di-Photonic Decays}

The annihilation decays of charmonium resonance for $S$ - wave $(n^{1}S_{0})$  and $P$ - wave $(n^{3}P_{0}\ \text{and}\ n^{3}P_{2})$ into two photons $(\gamma\gamma)$ with first order QCD correction factors are given by \cite{65,66}

\begin{equation}
	\begin{split}
		\Gamma\left(n^{1}S_{0} \rightarrow \gamma\gamma\right) &= \frac{2^{2}3\alpha^{2}e_{q}^{4}}{M(n^{1}S_{0})^{2}}|R_{nS}(0)|^{2}\left(1-\frac{3.4\alpha_{s}(\mu)}{\pi}\right) \,,
		\\
		\Gamma\left(n^{3}P_{0} \rightarrow \gamma\gamma\right) &= \frac{2^{4}27\alpha^{2}e_{q}^{4}}{M(n^{3}P_{0})^{4}}|R'_{nP}(0)|^{2}\left(1+\frac{0.2\alpha_{s}(\mu)}{\pi}\right) \,, 
		\\
		\Gamma\left(n^{3}P_{2} \rightarrow \gamma\gamma\right) &= \frac{2^{4}36\alpha^{2}e_{q}^{4}}{5M(n^{3}P_{2})^{4}}|R'_{nP}(0)|^{2}\left(1-\frac{16\alpha_{s}(\mu)}{3\pi}\right) \,.
	\end{split}
\end{equation}

\subsection{\label{sec:Tri-Photonic Decays} Tri-Photonic Decays}

The annihilation decays of charmonium resonance for $S$ - wave $(n^{3}S_{1})$  into three photons $(\gamma\gamma\gamma)$ with first order QCD correction factors are given by \cite{65,66}

\begin{eqnarray}
	\Gamma\left(n^{3}S_{1} \rightarrow \gamma\gamma\gamma\right) = && \frac{2^{2}4(\pi^{2}-9)\alpha^{3}e_{q}^{6}}{3\pi M(n^{3}S_{1})^{2}}|R_{nS}(0)|^{2}
	\nonumber\\
	&&\times \left(1-\frac{12.6\alpha_{s}(\mu)}{\pi}\right)\,.
\end{eqnarray}

\subsection{\label{sec:Di-Gluonic Decays} Di-Gluonic Decays}

The annihilation decays of charmonium resonance for $S$ - wave $(n^{1}S_{0})$, $P$ - wave $(n^{3}P_{0}\ \text{and}\ n^{3}P_{2})$, $D$ - wave $(n^{1}D_{2})$, $F$ - wave $(n^{3}F_{2} ,n^{3}F_{3}\ \text{and}\ n^{3}F_{4})$ and $G$ - wave $(n^{1}G_{4})$ into two gluons $(gg)$ with first order QCD correction factors are given by \cite{65,66,68}

\begin{equation}
	\begin{split}
		\Gamma\left(n^{1}S_{0} \rightarrow gg\right) &= \frac{2^{2}2\alpha_{s}^{2}(\mu)}{3M(n^{1}S_{0})^{2}}|R_{nS}(0)|^{2}\left(1+\frac{4.8\alpha_{s}(\mu)}{\pi}\right) \,, 
		\\
		\Gamma\left(n^{3}P_{0} \rightarrow gg\right) &= \frac{2^{4}6\alpha_{s}^{2}(\mu)}{M(n^{3}P_{0})^{4}}|R'_{nP}(0)|^{2}\left(1+\frac{9.5\alpha_{s}(\mu)}{\pi}\right) \,,
		\\
		\Gamma\left(n^{3}P_{2} \rightarrow gg\right) &= \frac{2^{4}8\alpha_{s}^{2}(\mu)}{5M(n^{3}P_{2})^{4}}|R'_{nP}(0)|^{2}\left(1-\frac{2.2\alpha_{s}(\mu)}{\pi}\right) \,,
		\\
		\Gamma\left(n^{1}D_{2} \rightarrow gg\right) &= \frac{2^{6}2\alpha_{s}^{2}(\mu)}{3\pi M(n^{1}D_{2})^{6}}|R''_{nD}(0)|^{2} \,,
		\\
		\Gamma\left(n^{3}F_{2} \rightarrow gg\right) &= \frac{2^{8}919\alpha_{s}^{2}(\mu)}{135 M(n^{3}F_{2})^{8}}|R'''_{nF}(0)|^{2} \,,
		\\
		\Gamma\left(n^{3}F_{3} \rightarrow gg\right) &= \frac{2^{8}20\alpha_{s}^{2}(\mu)}{27 M(n^{3}F_{3})^{8}}|R'''_{nF}(0)|^{2} \,,
		\\
		\Gamma\left(n^{3}F_{4} \rightarrow gg\right) &= \frac{2^{8}20\alpha_{s}^{2}(\mu)}{27 M(n^{3}F_{4})^{8}}|R'''_{nF}(0)|^{2} \,,
		\\
		\Gamma\left(n^{1}G_{4} \rightarrow gg\right) &= \frac{2^{10}2\alpha_{s}^{2}(\mu)}{3\pi M(n^{1}G_{4})^{10}}|R''''_{nG}(0)|^{2} \,.
	\end{split}
\end{equation}

\subsection{\label{sec:Tri-Gluonic Decays} Tri-Gluonic Decays}

The annihilation decays of charmonium resonance for $S$ - wave $(n^{3}S_{1})$, $P$ - wave $(n^{1}P_{1})$ and $D$ - wave $(n^{3}D_{1}, n^{3}D_{2}$ and $n^{3}D_{3})$ into three gluons $(ggg)$ with first order QCD correction factors are given by \cite{65,66,69}

\begin{align}
	\Gamma\left(n^{3}S_{1} \rightarrow ggg\right) &=\frac{2^{2}10(\pi^{2}-9)\alpha_{s}^{3}(\mu)}{81\pi M(n^{3}S_{1})^{2}}|R_{nS}(0)|^{2} \nonumber \\
	&\times\left(1-\frac{3.7\alpha_{s}(\mu)}{\pi}\right), \nonumber
	\\
	\Gamma\left(n^{1}P_{1} \rightarrow ggg\right) &= \frac{2^{4}20\alpha_{s}^{3}(\mu)}{9\pi M(n^{1}P_{1})^{4}}|R'_{nP}(0)|^{2}\ln\left(m_{q}\langle r\rangle \right), \nonumber
	\\
	\Gamma\left(n^{3}D_{1} \rightarrow ggg\right) &= \frac{2^{6}760\alpha_{s}^{3}(\mu)}{81\pi M(n^{3}D_{1})^{6}}|R''_{nD}(0)|^{2}\ln\left(4m_{q}\langle r\rangle \right), \nonumber
	\\
	\Gamma\left(n^{3}D_{2} \rightarrow ggg\right) &= \frac{2^{6}10\alpha_{s}^{3}(\mu)}{9\pi M(n^{3}D_{2})^{6}}|R''_{nD}(0)|^{2}\ln\left(4m_{q}\langle r\rangle \right), \nonumber
	\\
	\Gamma\left(n^{3}D_{3} \rightarrow ggg\right) &= \frac{2^{6}40\alpha_{s}^{3}(\mu)}{9\pi M(n^{3}D_{3})^{6}}|R''_{nD}(0)|^{2}\ln\left(4m_{q}\langle r\rangle \right)\,.
\end{align}

\subsection{\label{sec:Mixed Annihilation Decays} Mixed Annihilation Decays}

The annihilation decays of charmonium resonance $S$ - wave $(n^{3}S_{1})$ via strong and electromagnetic interaction, into a photon and two gluons $(\gamma gg)$ are given by \cite{54,64,65}
\begin{align}
	\Gamma\left(n^{3}S_{1} \rightarrow \gamma gg\right) &=\frac{2^{2}8(\pi^{2}-9)e_{q}^{2}\alpha\alpha_{s}^{3}(\mu)}{9\pi M(n^{3}S_{1})^{2}}|R_{nS}(0)|^{2} 
	\nonumber \\
	&\times\left(1-\frac{6.7\alpha_{s}(\mu)}{\pi}\right)\,.
\end{align}
The annihilation decays of charmonium resonance $P$ - wave $(n^{3}P_{1})$ into a light flavour meson and a gluon $(q\bar{q}g)$ are given by \cite{65,66}
\begin{equation}
	\Gamma\left(n^{3}P_{1} \rightarrow q\bar{q}g\right) = \frac{2^{4}8n_{f}\alpha_{s}^{3}(\mu)}{9\pi M(n^{3}P_{1})^{4}}|R'_{nP}(0)|^{2}\ln\left(m_{q}\langle r\rangle \right)\,,
\end{equation}
where $n_{f}$ is the number of flavors. 

\noindent For all the decays the strong coupling constatnt, $\alpha_{s}(\mu)$ is calculated using the expression
\begin{equation}
	\alpha_{s}(\mu) = \frac{4\pi}{\beta_{0}\ln\frac{\mu^{2}}{\Lambda^{2}}}\left(1-\frac{\beta_{1}\ln\left(\ln\frac{\mu^{2}}{\Lambda^{2}}\right)}{\beta_{0}^{2}\ln\frac{\mu^{2}}{\Lambda^{2}}}\right) ,
\end{equation}
where $\beta_{0}=11-(2/3)n_{f}$, $\beta_{1}=102- (18/3)n_{f}$, $\Lambda$ is the QCD constant, and $\mu$ is the reduced mass. 
\noindent All the annihilation decay widths for $c\bar{c}$ bound system are presented in Tables \ref{tab:13}-\ref{tab:19}, respectively.

\section{\label{sec:S-D Mixing} S-D Mixing}

Mixing effects between charmonium states are possible if their masses are close enough and their wavefunctions overlap. This mixing can have observable effects in spectroscopy, such as mass shifts and changes in decay properties, and it is a crucial factor to consider when interpreting experimental results. The higher excited charmonium $S$ and $D$ states have comparable masses, which might cause mixing between them due to non-perturbative effects within the quark-antiquark bound state. The mixed states can be represented in terms of pure $|nS\rangle$ and $|n'D\rangle$ states as \cite{70}
\begin{equation}
	\begin{split}
		|\phi\rangle &= \cos\theta|nS\rangle + \sin\theta|n'D\rangle \,,
		\\
		|\phi'\rangle &= -\sin\theta|nS\rangle + \cos\theta|n'D\rangle ,
	\end{split} 
\end{equation}
where $|\phi\rangle$ and $|\phi'\rangle$ are the mixed states, and $\theta$ is the mixing angle. The masses of the mixed states can be calculated by \cite{70}
\begin{align}
		M_{\phi} &= \left(\left(\frac{M_{nS}+M_{n'D}}{2}\right)+\left(\frac{M_{nS}-M_{n'D}}{2\cos2\theta}\right)\right) \,,
		\nonumber \\
		M_{\phi'} &= \left(\left(\frac{M_{nS}+M_{n'D}}{2}\right)+\left(\frac{M_{n'D}-M_{nS}}{2\cos2\theta}\right)\right) \,.
\end{align}
Here $M_{\phi}$ and $M_{\phi'}$ are the masses of the mixed states, and $M_{nS}$ and $M_{n'D}$ are the masses of the corresponding pure $S$ and $D$ states.

\noindent The leptonic decay widths of the mixed states are given by \cite{70,71}
\begin{widetext}
	\begin{equation}
		\begin{split}
			\Gamma_{\phi} &= \left(1-\frac{16\alpha_{s}(\mu)}{3\pi}\right)\left(\frac{2\alpha e_{q}}{M_{nS}}|R_{nS}(0)|\cos\theta +\frac{5\alpha e_{q}}{\sqrt{2}m_{q}^{2}M_{n'D}}|R''_{n'D}(0)|\sin\theta\right)^{2} \,,
			\\
			\Gamma_{\phi'} &= \left(1-\frac{16\alpha_{s}(\mu)}{3\pi}\right)\left(\frac{5\alpha e_{q}}{\sqrt{2}m_{q}^{2}M_{n'D}}|R''_{n'D}(0)|\cos\theta -\frac{2\alpha e_{q}}{M_{nS}}|R_{nS}(0)|\sin\theta\right)^{2}\,.
		\end{split}
	\end{equation}
\end{widetext}
The leptonic decays of the mixed states are fitted to the experimental data to obtain the mixing angle, which is then used to calculate the masses of the mixed states. Our results of $S-D$ mixing are presented in Table \ref{tab:20}.

\section{\label{sec:Results and Discussion} Results and Discussion}

In this study, a screened potential model within a relativistic framework is employed to compute the spectra and decay widths of $c\bar{c}$ bound system.The masses of $S$ - wave states are presented in Table \ref{tab:2} and are compared with the experimental data and other theoretical models. The masses of $1S$ and $2S$ charmonium are experimentally well established and our evaluation of the masses for $1S$ states, $\eta_{c}(1S) = 2986.3$ MeV and $J/\psi(1S) = 3094.1$ MeV, and for $2S$ states, $\eta_{c}(2S) = 3633.1$ MeV and $\psi(2S) = 3690.0$ MeV, are consistent with the experimental data \cite{exp}. The experimental values of hyperfine mass splitting between the singlet and triplet of $1S$ and $2S$ charmonium are $\Delta m(1S) = 113.3 \pm 0.7$ MeV and $\Delta m(2S) = 46.7 \pm 1.3$ MeV, respectively \cite{exp}. In our model, the hyperfine mass splitting given by $\Delta m(nS) = m[\psi(nS)] - m[\eta(nS)]$, are evaluated to be $\Delta m(1S) = 110.2$ MeV and $\Delta m(2S) = 56.9$ MeV. The experimental value for the mass difference between $m[\psi(2S)] - m[J/\psi(1S)] = 589.188 \pm 0.028$ MeV \cite{exp}, is evaluated to be $595.9$ MeV in our model. In our model the mass of $\eta_{c}(3S)$ charmonium state is evaluated to be $4011.9$ MeV. The $X(3940)$ can be a candidate for the $\eta_{c}(3S)$ since it is only observed in decay via $D^{*}\bar{D}$ channel with $J^{P} = 1^{+}$ and due to charge conjugate invariance \cite{102}. Our evaluted mass is higher than the experimental value of $3942_{-6}^{+7} \pm6$ MeV by $69.9$ MeV. In most potential models, the predicted mass of $\eta_{c}(3S)$ is usually higher than that of the $X(3940)$ \cite{102}. We assign the $\psi(4040)$ with mass $4048.2$ MeV to be the $\psi(3S)$ charmonium state. Generally, $\psi(4415)$ is considered to be $\psi(4S)$ charmonium state \cite{72}, with other predictions such as $\psi_{1}(3D)$ \cite{73}, $\psi(5S)$ \cite{55}, and mixture of $\psi(4S) - \psi(3D)$ \cite{70}. The masses of $\psi(4S)$ and $\psi(5S)$ charmonium state in our model are evaluated to be $4294.4$ MeV and $4468.2$ MeV. In our model the $\psi(5S)$ charmonium state comes out to be a better assignment for $\psi(4415)$. $\psi(4230)$ also known as $Y(4230/4260)$, has many interpretations in the literature such as hybrid mesons $(c\bar{c}g)$ \cite{74}, hadronic molecules $(q\bar{c})(\bar{q}c)$ \cite{75}, $\psi(4S)$ charmonium state \cite{55}, $\psi(3D)$ charmonium state \cite{76}, tetraquark $cs\bar{c}\bar{s}$ \cite{77}, etc. In our model $\psi(4230)$ is a better candidate for $\psi(4S)$ charmonium state compared to $\psi(4415)$. Recently observed $Y(4500)$ has been suggested as a mixture of $\psi(5S) - \psi(4D)$ charmonium states \cite{78}. In our model $Y(4500)$  is better suited as $\psi(5S)$ charmonium state in terms of conventional $(q\bar{q})$ structure, with mass value of $4468.2$ MeV , which is in agreement with the experimental value \cite{34}. There is a clear overlap between the conventional $c\bar{c}$ states and  $\psi(4230)$, $\psi(4415)$ and $Y(4500)$ through $S-D$ mixing in our model.
In the final part of this dicussion, we discuss the $S-D$ mixing for  $\psi(2S), \psi(4040), \psi(4230), \psi(4415)$, and $Y(4500)$ and our final assignments for these states are presented.

\begin{table}
	\caption{\label{tab:2} $S$ wave mass spectrum of $c\bar{c}$ bound system in MeV}
	\begin{ruledtabular}
	\begin{tabular}{cccccccc}
		States & Ours & Exp\cite{exp} & \cite{72} & \cite{79} & \cite{80} & \cite{81} & \cite{82}
		\\
		\hline 
		$1^{1}S_{0}$ & 2986.3 & 2983.9 & 2989 & 2995 & 2977.8 & 2991.95 & 2981
		\\
		$2^{1}S_{0}$ & 3633.1 & 3637.5 & 3572 & 3606 & 3630.5 & 3625.3 & 3635
		\\
		$3^{1}S_{0}$ & 4011.9 &  & 3998 & 4000 & 3990.8 & 4029.4 & 3989
		\\
		$4^{1}S_{0}$ & 4269.3 &  & 4372 & 4328 & 4262.1 & 4059.4 & 4401
		\\
		$5^{1}S_{0}$ & 4450.8 &  & 4714 & 4622 & 4439.2 & 4639.3 & 4811
		\\
		$6^{1}S_{0}$ & 4577.7 &  & 5033 & 4893 &  &  & 5155
		\\
		\hline
		$1^{3}S_{1}$ & 3094.1 & 3096.9 & 3094 & 3094 & 3096.7 & 3094.3 & 3096
		\\
		$2^{3}S_{1}$ & 3690.0 & 3686.1 & 3649 & 3649 & 3684.4 & 3667.6 & 3685
		\\
		$3^{3}S_{1}$ & 4048.2 & 4039.6 & 4062 & 4036 & 4022.4 & 4059.4 & 4039
		\\
		$4^{3}S_{1}$ & 4294.4 & 4415.0 & 4428 & 4362 & 4266.4 & 4356.2 & 4427
		\\
		$5^{3}S_{1}$ & 4468.2 &  & 4763 & 4654 & 4441.5 & 4660.7 & 4837
		\\
		$6^{3}S_{1}$ & 4589.4 &  & 5075 & 4925 &  &  & 5167
	\end{tabular}
	\end{ruledtabular}
\end{table}

\begin{table}
	\caption{\label{tab:3} $P$ wave mass spectrum of $c\bar{c}$ bound system in MeV}
	\begin{ruledtabular}
	\begin{tabular}{cccccccc}
		States & Ours & Exp \cite{exp} & \cite{72} & \cite{79} & \cite{80} & \cite{81} & \cite{82}
		\\
		\hline
		$1^{1}P_{1}$ & 3517.2 & 3525.4 & 3527 & 3534 & 3518.7 & 3544.2 & 3525 
		\\
		$2^{1}P_{1}$ & 3927.6 &  & 3975 & 3936 & 3956.2 & 3950.7 & 3926 
		\\
		$3^{1}P_{1}$ & 4209.9 &  & 4364 & 4269 & 4231.1 & 4283.0 & 4337 
		\\
		$4^{1}P_{1}$ & 4413.7 &  & 4716 &  & 4446.4 & 4571.4 & 4744 
		\\
		$5^{1}P_{1}$ & 4561.4 &  &  &  &  & 4829.8 &  
		\\
		\hline
		$1^{3}P_{0}$ & 3434.4 & 3414.7 & 3473 & 3457 & 3418.4 & 3472.2 & 3413 
		\\
		$2^{3}P_{0}$ & 3857.5 & 3862.0 & 3918 & 3866 & 3824.9 & 3884.7 & 3870 
		\\
		$3^{3}P_{0}$ & 4151.5 &  & 4306 & 4197 & 4136.0 & 4218.6 & 4301 
		\\
		$4^{3}P_{0}$ & 4366.6 &  & 4659 &  & 4383.2 & 4507.6 & 4698 
		\\
		$5^{3}P_{0}$ & 4525.1 &  &  &  &  & 4765.9 &  
		\\
		\hline
		$1^{3}P_{1}$ & 3514.2 & 3510.7 & 3506 & 3523 & 3513.0 & 3543.1 & 3511 
		\\
		$2^{3}P_{1}$ & 3924.7 & 3871.6 & 3949 & 3925 & 3901.8 & 3950.6 & 3906 
		\\
		$3^{3}P_{1}$ & 4206.9 &  & 4336 & 4257 & 4174.6 & 4282.5 & 4319 
		\\
		$4^{3}P_{1}$ & 4410.9 &  & 4688 &  & 4409.1 & 4570.2 & 4728 
		\\
		$5^{3}P_{1}$ & 4559.0 &  &  &  &  & 4827.8 & 
		\\
		\hline
		$1^{3}P_{2}$ & 3556.2 & 3556.2 & 3551 & 3556 & 3554.2 & 3584.0 & 3555
		\\
		$2^{3}P_{2}$ & 3965.2 & 3922.5 & 4002 & 3956 & 3921.2 & 3993.8 & 3949
		\\
		$3^{3}P_{2}$ & 4242.7 &  & 4392 & 4290 & 4203.7 & 4327.5 & 4354
		\\
		$4^{3}P_{2}$ & 4440.6 &  & 4744 &  & 4415.1 & 4616.9 & 4763
		\\
		$5^{3}P_{2}$ & 4582.3 &  &  &  &  & 4875.9 & 
	\end{tabular}
	\end{ruledtabular}
\end{table}

The masses of $P$ - wave states are presented in Table \ref{tab:3} and are compared with experimental data and other theoretical models. All the $1P$ states of charmonium have been experimentally observed and our masses for $1P$ states, $h_{c}(1P) = 3517.2$ MeV, $\chi_{c0}(1P) = 3434.4$ MeV, $\chi_{c1}(1P) = 3514.2$ MeV and $\chi_{c2}(1P) = 3556.2$ MeV are consistent with the measured values \cite{exp}. The experimental values of fine splitting for $1P$ charmonium state are $m[\chi_{c2}(1P)] - m[\chi_{c1}(1P)] = 45.5 \pm 0.2$ MeV and $m[\chi_{c1}(1P)] - m[\chi_{c0}(1P)] = 95.9 \pm 0.4$ MeV \cite{exp}, which comes out to be $42$ MeV and $79.8$ MeV, respectively in our model. $\chi_{c0}(3860)$ is assigned to $\chi_{c0}(2P)$ charmonium state with the mass of $3857.5$ MeV, which is in excellent agreement with the experimentally measured value \cite{exp}. The $X(3872)$ is a candidate for the $\chi_{c1}(2P)$ charmonium state. The mass of $\chi_{c1}(2P)$ in our work is evaluated to $3924.7$ MeV which is $53.1$ MeV higher than the observed mass of $X(3872)$. This suppression of the mass value is attributed to the coupling effects of the states due proximity with $D^{0}\bar{D}^{*0}$ threshold. In context of pure charmonium state, $\chi_{c0}(3915)$ is predicted to be $\chi_{c0}(2P)$ state \cite{72}. The $\chi_{c0}(3915)$ has many interpretations in the literature such as $D_{s}\bar{D}_{s}$ molecule \cite{83}, a tetraquark \cite{84}, etc. In our model  $\chi_{c0}(3915)$ cannot be assigned to $\chi_{c0}(2P)$ state. Further experimental data is required in order to make a more accurate determination. The $\chi_{c2}(3930)$ is assigned to $\chi_{c2}(2P)$ charmonium state in our model, with mass of $3965.2$ MeV and in agreement with experimental data \cite{exp}. $X(4140)$ and $X(4274)$ both are considered as possible candidates for $\chi_{c1}(3P)$ charmonium state \cite{85,86,87}. Other interpretations of $X(4140)$ and $X(4274)$ include a $cs\bar{c}\bar{s}$ tetraquark state \cite{88,89}, as axial vector tetraquark state \cite{90}, etc. In our model the mass of $\chi_{c1}(3P)$ charmonium state is calculated to be $4206.9$ MeV which is higher by $60.1$ MeV and lower by $66.1$ MeV compared to the experimental masses of $X(4140)$ and $X(4274)$, respectively. $X(4350)$ is interpreted as $\chi_{c2}(3P)$ charmonium state \cite{91}, molecular state \cite{92,93}, mixing of $c\bar{c}-D_{s}^{*}\bar{D}_{s}^{*}$ \cite{94}, etc. In our model $X(4350)$ is assigned to $\chi_{c0}(4P)$ charmonium state with mass of $4366.6$ MeV and $J^{PC} = 0^{++}$, which agrees with the experimental mass \cite{exp}. $X(4500)$ and $X(4700)$ show different properties compared to the standard $q\bar{q}$ structure and are considered exotic structures. $X(4500)$ and $X(4700)$ are commonly assumed to be a teraquark structures \cite{88,89}. In ref. \cite{87}, these structures are investigated with re-scattering effect and concluded that $X(4500)$ can be a genuine charmonium resonance. In our model $X(4500)$ can be assigned to $\chi_{c0}(5P)$ charmonium state with mass of $4525.1$ MeV, consistent with the experimental mass and $J^{PC} = 0^{++}$ \cite{exp}.

\begin{table}
	\caption{\label{tab:4} $D$ wave mass spectrum of $c\bar{c}$ bound system in MeV}
	\begin{ruledtabular}
	\begin{tabular}{cccccccc}
		States & Ours & Exp\cite{exp} & \cite{72} & \cite{79} & \cite{80} & \cite{81} & \cite{82}
		\\
		\hline 
		$1^{1}D_{2}$ & 3826.3 &  & 3780 & 3802 & 3662.2 & 3837.2 & 3807
		\\
		$2^{1}D_{2}$ & 4135.3 &  & 4203 & 4150 & 3968.9 & 4182.5 & 4196
		\\
		$3^{1}D_{2}$ & 4357.9 &  & 4566 & 4455 & 4236.2 & 4480.2 & 4549
		\\
		$4^{1}D_{2}$ & 4521.2 &  & 4901 &  & 4477.6 & 4745.6 & 4898
		\\
		$5^{1}D_{2}$ & 4639.9 &  &  &  & 4699.9 & 4987.0 & 
		\\
		\hline
		$1^{3}D_{1}$ & 3817.1 & 3773.7 & 3785 & 3799 & 3745.3 & 3830.2 & 3783
		\\
		$2^{3}D_{1}$ & 4123.3 & 4191.0 & 4196 & 4145 & 4038.9 & 4173.7 & 4150
		\\
		$3^{3}D_{1}$ & 4346.0 &  & 4562 & 4448 & 4300.6 & 4470.4 & 4507
		\\
		$4^{3}D_{1}$ & 4510.8 &  & 4898 &  & 4533.6 & 4735.1 & 4857
		\\
		$5^{3}D_{1}$ & 4631.5 &  &  &  & 4751.6 & 4976.0 & 
		\\
		\hline
		$1^{3}D_{2}$ & 3830.3 & 3823.7 & 3800 & 3805 & 3756.1 & 3841.1 & 3795
		\\
		$2^{3}D_{2}$ & 4137.5 &  & 4203 & 4152 & 4048.4 & 4186.7 & 4190
		\\
		$3^{3}D_{2}$ & 4359.2 &  & 4566 & 4456 & 4307.0 & 4484.6 & 4544
		\\
		$4^{3}D_{2}$ & 4522.1 &  & 4901 &  & 4541.6 & 4750.2 & 4896
		\\
		$5^{3}D_{2}$ & 4640.5 &  &  &  & 4758.9 & 4991.8 & 
		\\
		\hline
		$1^{3}D_{3}$ & 3829.3 & 3842.7 & 3806 & 3801 & 3769.6 & 3844.1 & 3813
		\\
		$2^{3}D_{3}$ & 4141.8 &  & 4206 & 4151 & 4060.7 & 4195.2 & 4220
		\\
		$3^{3}D_{3}$ & 4365.3 &  & 4568 & 4457 & 4317.2 & 4497.1 & 4574
		\\
		$4^{3}D_{3}$ & 4528.2 &  & 4902 &  & 4552.0 & 4765.6 & 4920
		\\
		$5^{3}D_{3}$ & 4645.8 &  &  &  & 4768.7 & 5009.7 & 
	\end{tabular}
	\end{ruledtabular}
\end{table}

The masses of $D$ - wave states are presented in Table \ref{tab:4} and compared with the experimental data and other theoretical models. Experimentally all the $1D$ states of charmonium have yet to be discovered. The $\psi(3770)$ is attributed to $\psi_{1}(1D)$ charmonium state. In our evaluation the mass of $\psi_{1}(1D)$ is $3817.1$ MeV, which is higher by $43.4$ MeV compared to experimental value. This difference in mass is caused due to mixture of $\psi(2S)-\psi(1D)$ charmonium states as suggested in ref. \cite{70,95}. The $\psi_{2}(3823)$ is considered to be the $\psi_{2}(1D)$ charmonium state in the literature. The mass computed for $\psi_{2}(1D)$ in our model is $3830.3$ MeV, which agrees with the experimental value \cite{exp}. The experimental value for the mass difference between $m[\psi_{2}(3823)]-m[\psi(2S)] = 137.98\pm 0.53 \pm 0.14$ MeV \cite{exp}, and in our model it is evaluted to be $140.3$ MeV. The recently discovered $\psi(3842)$ is widely considered to be the $\psi_{3}(1D)$ charmonium state. In our model the mass of $\psi_{3}(1D)$ state is $3829.3$ MeV in agreement with the experimental value \cite{exp}. The $\psi(4160)$ is conventionally considered to be $\psi_{1}(2D)$ charmonium state and in our evaluation of the mass for $\psi_{1}(2D)$ is obtained to be $4123.3$ MeV which is lower by $67.7$ MeV compared to experimental value \cite{exp}. It is suggested that the $S-D$ mixing can be very prominent in $\psi(4160)$ \cite{55,70,96}. In literature the $\psi(4360)$ is predicted to be hadrocharmonium \cite{97}, $\psi(4S)$ charmonium state \cite{73}, $\psi_{1}(3D)$ charmonium state \cite{55,98}, mixture of $\psi(4S)-\psi(3D)$ charmonium states \cite{70}, etc. In terms of conventional charmoium resonance, $\psi(4360)$ is a better candidate for $\psi_{1}(3D)$ charmonium state with mass of $4346.0$ MeV. $Y(4630)$ and $Y(4660)$ have very similar masses but their decay width and $J^{PC}$ are significantly different \cite{exp}. $Y(4630)$ and $Y(4660)$ are analysed as tetraquarks states \cite{99}, conventional charmonium states \cite{55,98}, etc. $Y(4630)$ cannot be accommodated to any of the conventional charmonium states in our model due to its mass and $J^{PC}$ value, but $Y(4660)$ can be assigned to $\psi_{1}(5D)$ charmonium state with mass of $4631.5$ MeV, which is in excellent agreement with the experimental value \cite{exp}. The $\psi(3770)$, $\psi(4160)$ and $\psi(4360)$ as mixture of $S-D$ states are also studied and presented later in this discussion.

\begin{table}
	\caption{\label{tab:5} $F$ and $G$ wave mass spectrum of $c\bar{c}$ bound system in MeV}
	\begin{ruledtabular}
	\begin{tabular}{cccccccc}
		States & Ours & \cite{72} & \cite{81} & States & Ours & \cite{81} & \cite{82}
		\\
		\hline 
		$1^{1}F_{3}$ & 4056.6 & 4039 & 4069.0 & $1^{1}G_{4}$ & 4241.2 & 4271.7 & 4345
		\\
		$2^{1}F_{3}$ & 4299.0 & 4413 & 4378.3 & $2^{1}G_{4}$ & 4434.3 &  & 
		\\
		$3^{1}F_{3}$ & 4477.5 & 4756 & 4652.6 & $3^{1}G_{4}$ & 4577.6 &  & 
		\\
		$4^{1}F_{3}$ & 4609.3 &  & 4901.1 & $4^{1}G_{4}$ & 4683.6 &  & 
		\\
		$5^{1}F_{3}$ & 4704.9 &  &  & $5^{1}G_{4}$ & 4760.0 &  & 
		\\
		\hline
		$1^{3}F_{2}$ & 4064.8 & 4015 & 4078.1 & $1^{3}G_{3}$ & 4254.8 & 4289.0 & 4321
		\\
		$2^{3}F_{2}$ & 4302.1 & 4403 & 4384.3 & $2^{3}G_{3}$ & 4442.7 &  & 
		\\
		$3^{3}F_{2}$ & 4478.1 & 4751 & 4656.5 & $3^{3}G_{3}$ & 4582.8 &  & 
		\\
		$4^{3}F_{2}$ & 4608.7 &  & 4903.6 & $4^{3}G_{3}$ & 4686.8 &  & 
		\\
		$5^{3}F_{2}$ & 4703.9 &  &  & $5^{3}G_{3}$ & 4761.9 &  & 
		\\
		\hline
		$1^{3}F_{3}$ & 4061.2 & 4039 & 4073.5 & $1^{3}G_{4}$ & 4245.2 & 4276.3 & 4343
		\\
		$2^{3}F_{3}$ & 4301.9 & 4413 & 4382.3 & $2^{3}G_{4}$ & 4442.7 &  & 
		\\
		$3^{3}F_{3}$ & 4479.4 & 4756 & 4656.3 & $3^{3}G_{4}$ & 4579.5 &  & 
		\\
		$4^{3}F_{3}$ & 4610.6 &  & 4904.8 & $4^{3}G_{4}$ & 4684.9 &  & 
		\\
		$5^{3}F_{3}$ & 4705.7 &  &  & $5^{3}G_{4}$ & 4760.8 &  & 
		\\
		\hline
		$1^{3}F_{4}$ & 4048.6 & 4052 & 4060.98 & $1^{3}G_{5}$ & 4229.2 & 4257.7 & 4357
		\\
		$2^{3}F_{4}$ & 4295.3 & 4418 & 4373.3 & $2^{3}G_{5}$ & 4426.7 &  & 
		\\
		$3^{3}F_{4}$ & 4476.0 & 4759 & 4649.9 & $3^{3}G_{5}$ & 4572.8 &  & 
		\\
		$4^{3}F_{4}$ & 4609.0 &  & 4900.2 & $4^{3}G_{5}$ & 4680.6 &  & 
		\\
		$5^{3}F_{4}$ & 4705.1 &  &  & $5^{3}G_{5}$ & 4758.2 &  & 
	\end{tabular}
	\end{ruledtabular}
\end{table}

The masses of $F$ and $G$ - wave states are presented in Table \ref{tab:5} and compared with other theoretical models. There is significant difference between the masses of excited $F$ and $G$ states when compared to other models.

Pseudoscalar $f_{P}$ and vector $f_{V}$ decay constants are presented in Table \ref{tab:6} and $f_{\chi_{0}}$ and $f_{\chi_{1}}$ decay constants are presented in Table \ref{tab:7}, with the results from other potential models. Our calculated values for the pseudoscalar $f_{P}$ and vector $f_{V}$ decay constants are higher compared to the experimental values and other theoretical models. In contrast to experimental data, the decay constants of pseudoscalar states have larger values than the decay constants of vector states, which is also observed in other potential models. Our results for $f_{\chi_{0}}$ and $f_{\chi_{1}}$ decay constants have decreasing trend compared to ref. \cite{60}

\begin{table}
	\caption{\label{tab:6} Psuedoscalar $f_{P}$ and vector $f_{V}$ decay constants of $c\bar{c}$ bound system in MeV}
	\begin{ruledtabular}
	\begin{tabular}{ccccccc}
		States & Ours & Exp\cite{exp} & \cite{60} & \cite{100} & \cite{79} & \cite{101}
		\\
		\hline 
		$1^{1}S_{0}$ & 558.7 & 335$\pm$75 & 519.0 & 350.314 & 395 & 402 
		\\
		$2^{1}S_{0}$ & 430.9 &  & 318.1 & 278.447 & 237 & 240 
		\\
		$3^{1}S_{0}$ & 361.4 &  & 267.5 & 249.253 & 208 & 193 
		\\
		$4^{1}S_{0}$ & 308.8 &  & 240.8 & 231.211 & 193 & 
		\\
		$5^{1}S_{0}$ & 262.9 &  & 223.2 & 218.241 & 184 & 
		\\
		$6^{1}S_{0}$ & 219.0 &  & 210.2 & 208.163 & 177 & 
		\\
		\hline
		$1^{3}S_{1}$ & 531.4 & 416$\pm$6 & 332.6 & 325.876 & 402 & 393 
		\\
		$2^{3}S_{1}$ & 414.0 & 304$\pm$4 & 238.0 & 257.340 & 239 & 293 
		\\
		$3^{3}S_{1}$ & 348.3 &  & 206.9 & 229.857 & 209 & 258 
		\\
		$4^{3}S_{1}$ & 298.1 &  & 189.0 & 212.959 & 194 &  
		\\
		$5^{3}S_{1}$ & 254.0 &  & 176.8 & 200.848 & 185 &  
		\\
		$6^{3}S_{1}$ & 211.8 &  & 167.6 & 191.459 & 177 &  
	\end{tabular}
	\end{ruledtabular}
\end{table}

\begin{table}
	\caption{\label{tab:7} $f_{\chi_{0}}$ and $f_{\chi_{1}}$ decay constants of $c\bar{c}$ bound system in MeV}
	\begin{ruledtabular}
	\begin{tabular}{cccccc}
		States & Ours & \cite{60} & States & Ours & \cite{60}
		\\
		\hline 
		$1^{3}P_{0}$ & 384.5 & 754.7 & $1^{3}P_{1}$ & 433.9 & 788.3 
		\\
		$2^{3}P_{0}$ & 395.9 & 841.5 & $2^{3}P_{1}$ & 449.3 & 913.2 
		\\
		$3^{3}P_{0}$ & 375.4 & 899.6 & $3^{3}P_{1}$ & 427.7 & 994.9 
		\\
		$4^{3}P_{0}$ & 343.1 & 949.1 & $4^{3}P_{1}$ & 392.2 & 1059.4 
		\\
		$5^{3}P_{0}$ & 303.4 & 992.1 & $5^{3}P_{1}$ & 347.7 & 1113.7 
	\end{tabular}
	\end{ruledtabular}
\end{table}

\begin{table*}
	\caption{\label{tab:8} $E1$ transition width $\Gamma_{E1}$ in keV and photon energy $E_{\gamma}$ in MeV of $S$ wave states for $c\bar{c}$ bound system}
	\begin{ruledtabular}
	\begin{tabular}{ccccccccc}
		Initial & Final & Ours & Ours  & Exp\cite{exp} & \cite{81} & \cite{53} & \cite{100} & \cite{102} 
		\\
		State & State & $E_{\gamma}$ & $\Gamma_{E1}$ &  &  &  &  & 
		\\
		\hline 
		$2^{1}S_{0}$ & $1^{1}P_{1}$ & 114.07 & 38.46 &  & 45.1543 & 36 & 36.197 & 56
		\\
		\hline
		$2^{3}S_{1}$ & $1^{3}P_{0}$ & 246.77 & 37.62 & 28.78$\pm$1.4 & 58.849 & 26 & 21.863 & 29
		\\
		& $1^{3}P_{1}$ & 171.66 & 41.52 & 28.66$\pm$1.5 & 50.9538 & 29 & 43.292 & 38
		\\
		& $1^{3}P_{2}$ & 131.42 & 32.33 & 27.99$\pm$1.3 & 35.503 & 24 & 62.312 & 39
		\\
		\hline
		$3^{1}S_{0}$ & $2^{1}P_{1}$ & 83.50 & 42.69 &  & 36.4002 & 64 & 51.917 & 
		\\
		& $1^{1}P_{1}$ & 464.28 & 16.96 &  & 2.9957 & 28 & 178.312 & 7.7
		\\
		\hline
		$3^{3}S_{1}$ & $2^{3}P_{0}$ & 186.60 & 44.79 &  & 28.7006 & 22 & 31.839 & 
		\\
		& $2^{3}P_{1}$ & 122.00 & 42.25 &  & 17.0589 & 43 & 64.234 & 
		\\
		& $2^{3}P_{2}$ & 82.50 & 22.90 &  & 4.044 & 48 & 86.472 & 
		\\
		& $1^{3}P_{0}$ & 567.62 & 7.61 &  & 0.1187 & 0.63 & 46.872 & 8.1
		\\
		& $1^{3}P_{1}$ & 499.15 & 9.54 &  & 0.2299 & 0.85 & 107.088 & 0.44
		\\
		& $1^{3}P_{2}$ & 462.48 & 9.16 &  & 0.3013 & 12.7 & 163.485 & 6.44
		\\
		\hline
		$4^{1}S_{0}$ & $3^{1}P_{1}$ & 59.02 & 29.69 &  & 48.9384 & 101 &  & 
		\\
		& $2^{1}P_{1}$ & 328.04 & 28.25 &  & 1.8977 & 31.3 &  & 
		\\
		& $1^{1}P_{1}$ & 685.86 & 7.55 &  & 1.6698 & 9.6 &  & 
		\\
		\hline
		$4^{3}S_{1}$ & $3^{3}P_{0}$ & 140.56 & 37.43 &  & 95.5685 & 25 &  & 
		\\
		& $3^{3}P_{1}$ & 86.59 & 29.90 &  & 91.3162 & 54 &  & 
		\\
		& $3^{3}P_{2}$ & 51.44 & 11.03 &  & 44.9699 & 66 &  & 
		\\
		& $2^{3}P_{0}$ & 414.70 & 12.19 &  & 0.0941 & 0.39 &  & 
		\\
		& $2^{3}P_{1}$ & 353.80 & 14.71 &  & 0.1970 & 0.92 &  & 
		\\
		& $2^{3}P_{2}$ & 316.65 & 12.71 &  & 0.2513 & 15 &  & 
		\\
		& $1^{3}P_{0}$ & 773.91 & 2.65 &  & 0.1226 & 0.13 &  & 
		\\
		& $1^{3}P_{1}$ & 709.36 & 3.53 &  & 0.2880 & 0.53 &  & 
		\\
		& $1^{3}P_{2}$ & 674.79 & 3.55 &  & 0.4225 & 5.2 &  & 
	\end{tabular}
	\end{ruledtabular}
\end{table*}

The $E1$ transition width for $S,P,D,F$ and $G$ - wave states are presented in Tables \ref{tab:8}-\ref{tab:11}, respectively, and are compared with the experimental data and other theoretical models. The $S$ wave $E1$ transition widths in Table \ref{tab:8} are calculated for $\Gamma(2S \rightarrow \gamma P)$, $\Gamma(3S \rightarrow \gamma P)$ and $\Gamma(4S \rightarrow \gamma P)$. The $\Gamma(2S \rightarrow \gamma \chi_{c}(1P))$ results in our model are in good agreement with experimental data. $X(3940)$ which is assigned to the $\eta_{c}(3S)$ state in our model, has consistent transitions width for $\Gamma(\eta_{c}(3S) \rightarrow \gamma h_{c}(2P))$ across all the models but the $\Gamma(\eta_{c}(3S) \rightarrow \gamma h_{c}(1P))$ transitions width are significantly different. $\psi(4040)$ which is considered as the $\psi(3S)$ state, has transitions width of $\Gamma(\psi(3S) \rightarrow \gamma \chi_{c1}(2P)) < 272 \pm 34$ keV and $\Gamma(\psi(3S) \rightarrow \gamma \chi_{c2}(2P)) < 400 \pm 50$ keV \cite{exp}. Our evaluation of these decays are in agreement with the experimental observations.

The $P$ wave $E1$ transition widths in Table \ref{tab:9} are calculated for $\Gamma(1P \rightarrow \gamma S)$, $\Gamma(2P \rightarrow \gamma S)$, $\Gamma(2P \rightarrow \gamma D)$, $\Gamma(3P \rightarrow \gamma S)$ and $\Gamma(3P \rightarrow \gamma D)$. The $\Gamma(1P \rightarrow \gamma 1S)$ transition widths in our model are substantially better compared to other theoretical models and more consistent with the experimental results.The experimental values of transition widths for $\Gamma(X(3872) \rightarrow \gamma \psi(2S)) = 53.55 \pm 18.128$ keV and $\Gamma(X(3872) \rightarrow \gamma J/\psi(1S)) = 10.115 \pm 3.368$ keV and the ratio $\Gamma(X(3872) \rightarrow \gamma \psi(2S))/\Gamma(X(3872) \rightarrow \gamma J/\psi(1S)) = 2.6 \pm 0.6$ \cite{exp}. The $X(3872)$ is assigned to $\chi_{c1}(2P)$ charmonium state, and these transition widths are higher by a factor of $10$, with the ratio coming out to be $1.39$ in our model. This suggest that the $X(3872)$ has properties different than the conventional $q\bar{q}$ structure. $\chi_{0}(3860)$ and $\chi_{c2}(3930)$, are assigned to $\chi_{0}(2P)$ and $\chi_{c2}(2P)$ charmonium state in our model and their $E1$ transition widths are also calculated. $X(4140)$ and $X(4274)$ can be assigned to $\chi_{1}(3P)$ state and the transition widths for $\Gamma(\chi_{1}(3P) \rightarrow \gamma S)$ and $\Gamma(\chi_{1}(3P) \rightarrow \gamma D)$ are also estimated. Further experimental evidence is needed to make a firm prediction.

\begin{table*}
	\caption{\label{tab:9} $E1$ transition width $\Gamma_{E1}$ in keV and photon energy $E_{\gamma}$ in MeV of $P$ wave states for $c\bar{c}$ bound system}
	\begin{ruledtabular}
	\begin{tabular}{ccccccccc}
		Initial & Final & Ours & Ours  & Exp\cite{exp} & \cite{81} & \cite{53} & \cite{100} & \cite{102} 
		\\
		State & State & $E_{\gamma}$ & $\Gamma_{E1}$ &  &  &  &  & 
		\\
		\hline 
		$1^{1}P_{1}$ & $1^{1}S_{0}$ & 490.81 & 450.53 & 350$\pm$153.5 & 570.53 & 352 & 247.971 & 473 \\
		\hline
		$1^{3}P_{0}$ & $1^{3}S_{1}$ & 323.49 & 155.42 & 151.2$\pm$13.8 & 159.246 & 114 & 112.030 & 104 
		\\
		\hline
		$1^{3}P_{1}$ & $1^{3}S_{1}$ & 395.02 & 263.66 & 288$\pm$22.1 & 329.488 & 239 & 146.317 & 341 \\
		\hline
		$1^{3}P_{2}$ & $1^{3}S_{1}$ & 432.10 & 331.41 & 374.3$\pm$26.9 & 437.584 & 313 & 157.225 & 405 
		\\
		\hline
		$2^{1}P_{1}$ & $1^{1}D_{2}$ & 100.00 & 23.66 &  & 33.3607 & 27 &  & 51 
		\\
		& $2^{1}S_{0}$ & 283.43 & 244.02 &  & 349.823 & 218 & 163.646 & 274 
		\\
		& $1^{1}S_{0}$ & 828.48 & 157.92 &  & 114.078 & 85 & 329.384 & 116 
		\\
		\hline
		$2^{3}P_{0}$ & $1^{3}D_{1}$ & 40.19 & 1.61 &  & 33.2356 & 51 &  & 1.4 
		\\
		& $2^{3}S_{1}$ & 163.83 & 58.36 &  & 108.31 & 135 & 70.400 & 83 
		\\
		& $1^{3}S_{1}$ & 687.90 & 74.14 &  & 50.9651 & 1.3 & 173.324 & 28 
		\\
		\hline
		$2^{3}P_{1}$ & $1^{3}D_{1}$ & 106.15 & 7.03 &  & 31.154 & 21 &  & 11 
		\\
		& $1^{3}D_{2}$ & 93.29 & 14.51 &  & 21.7898 & 18 &  & 30 
		\\
		& $2^{3}S_{1}$ & 227.66 & 141.59 &  & 246.036 & 183 & 102.672 & 234 
		\\
		& $1^{3}S_{1}$ & 742.76 & 101.85 &  & 62.6288 & 14 & 210.958 & 33 
		\\
		\hline
		$2^{3}P_{2}$ & $1^{3}D_{1}$ & 145.39 & 0.69 &  & 2.3056 & 1 &  & 0.64 
		\\
		& $1^{3}D_{2}$ & 132.66 & 7.98 &  & 11.4779 & 5.6 &  & 10 
		\\
		& $1^{3}D_{3}$ & 133.63 & 45.60 &  & 60.6771 & 29 &  & 76 
		\\
		& $2^{3}S_{1}$ & 265.67 & 209.00 &  & 377.091 & 207 & 116.325 & 264 
		\\
		& $1^{3}S_{1}$ & 775.49 & 121.67 &  & 70.9875 & 53 & 227.915 & 111 
		\\
		\hline
		$3^{1}P_{1}$ & $2^{1}D_{2}$ & 73.94 & 26.27 &  & 52.3298 & 48 &  & 158 
		\\
		& $1^{1}D_{2}$ & 366.11 & 7.29 &  & 0.0161 & 5.7 &  & 0.56 
		\\
		& $3^{1}S_{0}$ & 193.23 & 162.92 &  & 358.257 & 208 &  & 332 
		\\
		& $2^{1}S_{0}$ & 537.24 & 118.75 &  & 58.3516 & 43 &  & 90 
		\\
		& $1^{1}S_{0}$ & 1045.74 & 71.37 &  & 47.532 & 38 &  & 76 
		\\
		\hline
		$3^{3}P_{0}$ & $2^{3}D_{1}$ & 28.09 & 1.51 &  & 10.7876 & 35 &  & 11 
		\\
		& $1^{3}D_{1}$ & 320.93 & 26.64 &  & 0.0906 & 9.7 &  & 30 
		\\
		& $3^{3}S_{1}$ & 101.60 & 29.56 &  & 140.784 & 145 &  & 72 
		\\
		& $2^{3}S_{1}$ & 435.81 & 48.69 &  & 29.2243 & 0.045 &  & 19 
		\\
		& $1^{3}S_{1}$ & 922.77 & 38.54 &  & 21.6538 & 1.5 &  & 22 
		\\
		\hline
		$3^{3}P_{1}$ & $2^{3}D_{1}$ & 82.81 & 9.10 &  & 23.1234 & 15 &  & 34 
		\\
		& $2^{3}D_{2}$ & 68.86 & 16.03 &  & 32.9113 & 35 &  & 92 
		\\
		& $1^{3}D_{1}$ & 371.79 & 2.04 &  & 0.0326 & 0.39 &  & 0.00029 
		\\
		& $1^{3}D_{2}$ & 359.79 & 4.81 &  & 0.0652 & 4.6 &  & 1.8 
		\\
		& $3^{3}S_{1}$ & 155.36 & 94.31 &  & 339.862 & 181 &  & 272 
		\\
		& $2^{3}S_{1}$ & 485.16 & 77.27 &  & 39.6913 & 8.9 &  & 31 
		\\
		& $1^{3}S_{1}$ & 965.69 & 48.36 &  & 24.8275 & 2.2 &  & 11 
		\\
		\hline
		$3^{3}P_{2}$ & $2^{3}D_{1}$ & 117.69 & 0.98 &  & 2.29 & 0.77 &  & 2.1
		\\
		& $2^{3}D_{2}$ & 103.86 & 10.39 &  & 20.3685 & 9.9 &  & 34
		\\
		& $2^{3}D_{3}$ & 99.71 & 51.87 &  & 95.098 & 51 &  & 231
		\\
		& $1^{3}D_{1}$ & 404.24 & 0.15 &  & 0.0016 & 0.001 &  & 0.062
		\\
		& $1^{3}D_{2}$ & 392.35 & 1.79 &  & 0.017 & 0.13 &  & 0.42
		\\
		& $1^{3}D_{3}$ & 393.25 & 10.24 &  & 0.0939 & 6.8 &  & 1.4
		\\
		& $3^{3}S_{1}$ & 189.63 & 155.78 &  & 552.378 & 199 &  & 360
		\\
		& $2^{3}S_{1}$ & 516.65 & 100.99 &  & 48.2333 & 30 &  & 97
		\\
		& $1^{3}S_{1}$ & 993.14 & 55.67 &  & 27.2067 & 19 &  & 59
	\end{tabular}
	\end{ruledtabular}
\end{table*}

The $D$ wave $E1$ transition widths in Table \ref{tab:10} are calculated for $\Gamma(1D \rightarrow \gamma P)$, $\Gamma(2D \rightarrow \gamma P)$ and $\Gamma(2D \rightarrow \gamma F)$. Compared to the experimental results, the transition widths for $\Gamma(\psi_{1}(1D) \rightarrow \gamma \chi_{c}(1P))$ are higher in our calculations, except for $\Gamma(\psi_{1}(1D) \rightarrow \gamma \chi_{c2}(1P))$. This trend is also observed in other potential models. This variance could be due to the $S-D$ mixing in the $\psi(3770)$ state. The $E1$ transition widths are not experimentally measured for $\psi(3823)$ but the $\psi(3823) \rightarrow \gamma \chi_{c1}(1P)$ and $\psi(3823) \rightarrow \gamma \chi_{c2}(1P)$ are observed and their ratio is determined to be $\Gamma(\psi(3823) \rightarrow \gamma \chi_{c2}(1P))/\Gamma(\psi(3823) \rightarrow \gamma \chi_{c1}(1P)) = 0.28_{-0.11}^{+0.14} \pm 0.02$ \cite{exp}. $\psi(3823)$ which is assigned to $\psi_{2}(1D)$ state in our model, the transitions widths are calculated to be $285.63$ keV and $66.48$ keV and the ratio is evaluated to be $0.233$. $\psi(4160)$ has an experimentally measured transition width of $\psi(4160) \rightarrow \gamma X(3872) < 126 \pm 18$  keV \cite{exp}. In our model $\psi(4160)$ is considered to be the $\psi_{1}(2D)$ and the transition width is estimated to be $80.76$ keV.

\begin{table*}
	\caption{\label{tab:10} $E1$ transition width $\Gamma_{E1}$ in keV and photon energy $E_{\gamma}$ in MeV of $D$ wave states for $c\bar{c}$ bound system}
	\begin{ruledtabular}
	\begin{tabular}{ccccccccc}
		Initial & Final & Ours & Ours  & Exp\cite{exp} & \cite{81} & \cite{53} & \cite{100} & \cite{102} 
		\\
		State & State & $E_{\gamma}$ & $\Gamma_{E1}$ &  &  &  &  & 
		\\
		\hline 
		$1^{1}D_{2}$ & $1^{1}P_{1}$ & 296.60 & 360.19 &  & 480.201 & 344 & 205.93 & 374 
		\\
		\hline
		$1^{3}D_{1}$ & $1^{3}P_{0}$ & 363.50 & 332.89 & 187.7$\pm$23.2 & 394.563 & 213 & 161.504 & 362 
		\\
		& $1^{3}P_{1}$ & 290.89 & 142.65 & 67.7$\pm$8.7 & 122.833 & 77 & 93.775 & 153 
		\\
		& $1^{3}P_{2}$ & 251.99 & 6.49 & $<$17.4$\pm$0.64 & 4.7324 & 3.3 & 5.722 & 8.1 
		\\
		\hline
		$1^{3}D_{2}$ & $1^{3}P_{1}$ & 303.06 & 285.63 &  & 438.159 & 268 & 165.176 & 301 
		\\
		& $1^{3}P_{2}$ & 264.29 & 66.48 &  & 96.5234 & 66 & 50.317 & 82 
		\\
		\hline
		$1^{3}D_{3}$ & $1^{3}P_{2}$ & 263.36 & 263.43 &  & 397.684 & 296 & 175.212 & 302 
		\\
		\hline 
		$2^{1}D_{2}$ & $1^{1}F_{3}$ & 77.92 & 13.39 &  & 32.1154 & 23 &  &  
		\\
		& $2^{1}P_{1}$ & 202.46 & 216.68 &  & 363.001 & 296 &  &  
		\\
		& $1^{1}P_{1}$ & 571.88 & 121.55 &  & 35.3936 & 25 &  & 41 
		\\
		\hline
		$2^{3}D_{1}$ & $1^{3}F_{2}$ & 58.14 & 5.68 &  & 11.7495 & 17 &  &  
		\\
		& $2^{3}P_{0}$ & 257.24 & 213.87 &  & 339.764 & 191 &  &  
		\\
		& $2^{3}P_{1}$ & 193.81 & 80.76 &  & 116.186 & 114 &  &  
		\\
		& $2^{3}P_{2}$ & 155.02 & 2.98 &  & 3.9667 & 6.3 &  &  
		\\
		& $1^{3}P_{0}$ & 631.36 & 103.55 &  & 21.7083 & 35 &  & 109 
		\\
		& $1^{3}P_{1}$ & 564.13 & 47.69 &  & 11.4338 & 3.4 &  & 5.5 
		\\
		& $1^{3}P_{2}$ & 528.12 & 2.38 &  & 0.6305 & 0.027 &  & 0.080 
		\\
		\hline 
		$2^{3}D_{2}$ & $1^{3}F_{2}$ & 72.12 & 1.19 &  & 3.1069 & 2.4 &  &  
		\\
		& $1^{3}F_{3}$ & 75.59 & 10.89 &  & 28.0715 & 19 &  &  
		\\
		& $2^{3}P_{1}$ & 207.32 & 172.53 &  & 299.34 & 225 &  &  
		\\
		& $2^{3}P_{2}$ & 168.67 & 33.68 &  & 55.9308 & 65 &  &  
		\\
		& $1^{3}P_{1}$ & 576.37 & 94.34 &  & 22.9949 & 23 &  & 41 
		\\
		& $1^{3}P_{2}$ & 540.49 & 23.74 &  & 6.3982 & 0.62 &  & 1.5 
		\\
		\hline
		$2^{3}D_{3}$ & $1^{3}F_{2}$ & 76.30 & 0.03 &  & 0.0790 & 0.055 &  &  
		\\
		& $1^{3}F_{3}$ & 79.76 & 1.14 &  & 3.099 & 1.9 &  &  
		\\
		& $1^{3}F_{4}$ & 92.13 & 19.96 &  & 47.7232 & 26 &  &  
		\\
		& $2^{3}P_{2}$ & 172.75 & 143.58 &  & 253.36 & 272 &  &  
		\\
		& $1^{3}P_{2}$ & 544.18 & 97.87 &  & 26.5225 & 16 &  & 23 
	\end{tabular}
	\end{ruledtabular}
\end{table*}

\begin{table*}
	\caption{\label{tab:11} $E1$ transition width $\Gamma_{E1}$ in keV and photon energy $E_{\gamma}$ in MeV of $F$ and $G$ wave states for $c\bar{c}$ bound system}
	\begin{ruledtabular}
	\begin{tabular}{cccccc}
		Initial & Final & Ours & Ours & \cite{81} & \cite{53}   
		\\
		State & State & $E_{\gamma}$ & $\Gamma_{E1}$ &  &  
		\\
		\hline 
		$1^{1}F_{3}$ & $1^{1}D_{2}$ & 223.77 & 317.76 & 393.26 & 321 
		\\
		\hline
		$1^{3}F_{2}$ & $1^{3}D_{1}$ & 240.11 & 321.09 & 744.19 & 541 
		\\
		& $1^{3}D_{2}$ & 227.69 & 51.76 & 65.5198 & 69 
		\\
		& $1^{3}D_{3}$ & 228.64 & 1.49 & 1.7157 & 1.4 
		\\
		\hline
		$1^{3}F_{3}$ & $1^{3}D_{2}$ & 224.36 & 284.45 & 353.95 & 325 
		\\
		& $1^{3}D_{3}$ & 225.31 & 35.95 & 42.6351 & 38 
		\\
		\hline
		$1^{3}F_{4}$ & $1^{3}D_{3}$ & 213.35 & 279.88 & 326.749 & 334 
		\\
		\hline 
		$2^{1}F_{3}$ & $1^{1}G_{4}$ & 57.45 & 6.75 & 27.1812 & 20 
		\\
		& $2^{1}D_{2}$ & 160.63 & 193.58 & 363.36 & 284 
		\\
		& $1^{1}D_{2}$ & 446.74 & 99.17 & 17.4017 & 9.9 
		\\
		\hline
		$2^{3}F_{2}$ & $1^{3}G_{3}$ & 46.97 & 3.73 & 19.5170 & 16 
		\\
		& $2^{3}D_{1}$ & 175.05 & 202.56 & 461.5 & 295 
		\\
		& $2^{3}D_{2}$ & 161.42 & 30.49 & 58.601 & 50 
		\\
		& $2^{3}D_{3}$ & 157.32 & 0.81 & 1.4746 & 1.4 
		\\
		& $1^{3}D_{1}$ & 457.65 & 92.75 & 19.332 & 18 
		\\
		& $1^{3}D_{2}$ & 445.92 & 15.30 & 2.5806 & 0.39 
		\\
		& $1^{3}D_{3}$ & 446.81 & 0.44 & 0.0726 & 0.003 
		\\
		\hline
		$2^{3}F_{3}$ & $1^{3}G_{3}$ & 46.85 & 0.23 & 1.1459 & 1 
		\\
		& $1^{3}G_{4}$ & 56.43 & 6.00 & 25.0232 & 18 
		\\
		& $2^{3}D_{2}$ & 161.30 & 173.93 & 325.15 & 289 
		\\
		& $2^{3}D_{3}$ & 157.21 & 20.34 & 35.7473 & 35 
		\\
		& $1^{3}D_{2}$ & 445.81 & 87.34 & 14.599 & 11 
		\\
		& $1^{3}D_{3}$ & 446.69 & 11.01 & 1.7979 & 0.14
		\\
		\hline
		$2^{3}F_{4}$ & $1^{3}G_{3}$ & 40.25 & 0.001 & 0.0105 & 0.011 
		\\
		& $1^{3}G_{4}$ & 49.84 & 0.22 & 0.9993 & 0.84 
		\\
		& $1^{3}G_{5}$ & 65.56 & 9.43 & 32.7248 & 22 
		\\
		& $2^{3}D_{3}$ & 150.78 & 164.01 & 278.883 & 297 
		\\
		& $1^{3}D_{3}$ & 440.71 & 93.33 & 15.4547 & 8.6 
		\\
		\hline
		$1^{1}G_{4}$ & $1^{1}F_{3}$ & 180.57 & 282.23 & 406.979 & 374 
		\\
		\hline
		$1^{3}G_{3}$ & $1^{3}F_{2}$ & 185.85 & 279.88 & 419.428 & 366 
		\\
		& $1^{3}F_{3}$ & 189.22 & 25.68 & 39.0477 & 31 
		\\
		& $1^{3}F_{4}$ & 201.26 & 0.48 & 0.7286 & 0.52 
		\\
		\hline
		$1^{3}G_{4}$ & $1^{3}F_{3}$ & 179.95 & 262.19 & 382.394 & 349 
		\\
		& $1^{3}F_{4}$ & 192.02 & 20.76 & 30.2781 & 25 
		\\
		\hline
		$1^{3}G_{5}$ & $1^{3}F_{4}$ & 176.78 & 266.63 & 373.335 & 355 
	\end{tabular}
	\end{ruledtabular}
\end{table*}

The $F$ and $G$ wave $E1$ transitions widths in Table \ref{tab:11} are calculated for $\Gamma(1F \rightarrow \gamma D)$, $\Gamma(2F \rightarrow \gamma D)$, $\Gamma(2F \rightarrow \gamma G)$ and $\Gamma(1G \rightarrow \gamma F)$. The transition widths for $F$ and $G$ wave are compared with other potential models. Across all the models there is notable variation in the values.

The $S$ and $P$ wave $M1$ transition widths are presented in Table \ref{tab:12} and are compared with the available experimental data and other theoretical models. Our model show improvements in the $S$ wave estimates compared to other theoretical models. It is evident in the decay width values of $\Gamma(\psi(2S) \rightarrow \gamma \eta_{c}(1S))$ and $\Gamma(\eta_{c}(2S) \rightarrow \gamma J/\psi(1S))$. $M1$ transition widths of higher order $P$ states, $h_{c}(2P,3P)$ and $\chi_{c}(2P,3P)$ are also calculated and compared with other models. 

\begin{table}
	\caption{\label{tab:12} $M1$ transition width $\Gamma_{M1}$ in keV and photon energy $E_{\gamma}$ in MeV of $S$ and $P$ wave states for $c\bar{c}$ bound system}
	\begin{ruledtabular}
	\begin{tabular}{cccccccc}
		Initial & Final & Ours & Ours  & Exp\cite{exp} & \cite{81} & \cite{102} & \cite{53}  
		\\
		State & State & $E_{\gamma}$ & $\Gamma_{M1}$ &  &  &  &  
		\\
		\hline 
		$1^{3}S_{1}$ & $1^{1}S_{0}$ & 105.86 & 2.83 & 1.57$\pm$0.40 & 2.7654 & 2.2 & 2.4
		\\
		\hline
		$2^{1}S_{0}$ & $1^{3}S_{1}$ & 499.07 & 1.29 & 1.58$\pm$0.44 & 5.7919 & 6.9 & 5.6
		\\
		\hline
		$2^{3}S_{1}$ & $2^{1}S_{0}$ & 56.48 & 0.43 & 0.21$\pm$0.15 & 0.1980 & 3.8 & 0.17
		\\
		& $1^{1}S_{0}$ & 636.61 & 2.07 & 0.99$\pm$0.17 & 3.3698 & 0.096 & 9.6
		\\
		\hline
		$3^{1}S_{0}$ & $2^{3}S_{1}$ & 309.03 & 0.42 &  & 1.1424 & 1.6 & 0.84
		\\
		& $1^{3}S_{1}$ & 812.92 & 0.92 &  & 4.7885 & 6.5 & 6.9
		\\
		\hline
		$3^{3}S_{1}$ & $3^{1}S_{0}$ & 36.46 & 0.12 &  & 0.0023 & 0.044 & 0.067
		\\
		& $2^{1}S_{0}$ & 394.17 & 0.69 &  & 0.4702 & 0.71 & 2.6
		\\
		& $1^{1}S_{0}$ & 922.93 & 0.83 &  & 2.4944 & 3.7 & 9
		\\
		\hline
		$2^{1}P_{1}$ & $1^{3}P_{0}$ & 462.22 & 0.30 &  & 0.0269 & 5.3 & 1.5
		\\
		& $1^{3}P_{1}$ & 391.64 & 0.31 &  & 0.0476 & 0.13 & 0.36
		\\
		& $1^{3}P_{2}$ & 353.84 & 0.26 &  & 0.0589 & 1.2 & 0.11
		\\
		\hline
		$2^{3}P_{0}$ & $1^{1}P_{1}$ & 325.29 & 0.09 &  & 0.0258 & 4.8 & 0.5
		\\
		\hline
		$2^{3}P_{1}$ & $1^{1}P_{1}$ & 386.36 & 0.28 &  & 0.0412 & 0.13 & 0.045
		\\
		\hline
		$2^{3}P_{2}$ & $1^{1}P_{1}$ & 422.74 & 0.51 &  & 0.0537 & 0.89 & 1.3
		\\
		\hline
		$3^{1}P_{1}$ & $2^{3}P_{0}$ & 337.62 & 0.18 &  &  & 2.7 & 
		\\
		& $2^{3}P_{1}$ & 275.49 & 0.15 &  &  & 0.12 & 
		\\
		& $2^{3}P_{2}$ & 237.51 & 0.09 &  &  & 1.5 & 
		\\
		& $1^{3}P_{0}$ & 704.04 & 0.14 &  &  & 8.2 & 
		\\
		& $1^{3}P_{1}$ & 638.19 & 0.19 &  &  & 0.22 & 
		\\
		& $1^{3}P_{2}$ & 602.93 & 0.21 &  &  & 2.1 & 
		\\
		\hline
		$3^{3}P_{0}$ & $2^{1}P_{1}$ & 217.87 & 0.03 &  &  & 3.3 & 
		\\
		& $1^{1}P_{1}$ & 585.84 & 0.10 &  &  & 5.1 & 
		\\
		\hline
		$3^{3}P_{1}$ & $2^{1}P_{1}$ & 270.09 & 0.13 &  &  & 0.12 & 
		\\
		& $1^{1}P_{1}$ & 633.21 & 0.18 &  &  & 0.20 & 
		\\
		\hline
		$3^{3}P_{2}$ & $2^{1}P_{1}$ & 303.39 & 0.28 &  &  & 0.99 & 
		\\
		& $1^{1}P_{1}$ & 663.45 & 0.26 &  &  & 1.5 & 
	\end{tabular}
	\end{ruledtabular}
\end{table}

The di-leptonic decay widths $\Gamma(l^{+}l^{-})$ of charmonium states with $(\Gamma_{cf})$ and without $(\Gamma)$ the correction factor are evaluated for $\psi(nS)$ and $\psi_{1}(nD)$ in Table \ref{tab:13} and are compared with experimental data and other theoretical models. Di-leptonic decay widths for $\psi(1S)$ and $\psi(2S)$ are slightly higher by a magnitude of $1$ keV with respect to experimental data, but are consistent other models. Di-leptonic decay width of $\psi(4040)$ is measured to be $0.86 \pm 0.16$ keV \cite{exp}, which in our model is evaluated to be $2.17$ keV. The suppressed width can be due to an $S-D$ mixing component in the structure of $\psi(4040)$. $\psi(4415)$ which is conventionally considered to be $\psi(4S)$ charmonium state, has di-leptonic decay width of $0.58 \pm 0.07$ keV \cite{exp}. The $\psi(4415)$ which is considered to be the $\psi(5S)$ in terms of conventional charmonium state, has di-leptonic decay width of $1.05$ keV in our model. Our estimation is still higher compared to experimentally observed value. This implies that $\psi(4415)$ has a sizable $S-D$ mixing component which suppresses its di-leptonic decay width. The di-leptonic decay widths for $\psi(3770)$ and $\psi(4160)$ in our model are calculated to be $0.11$ keV and $0.18$ keV, respectively. These values are considerably smaller than the experimentally measured width. The assertion that $S-D$ mixing is very prominent in these states is further supported by this result. $Y(4360)$ and $Y(4660)$ are assigned as $\psi_{1}(3D)$ and $\psi_{1}(5D)$, respectively in our model. Their di-leptonic decay width are predicted to be $0.20$ keV and $0.16$ keV, respectively. The effect of $S-D$ mixing on the di-leptonic decay widths are discussed later in this discussion.

\begin{table}
	\caption{\label{tab:13} Di-leptonic decay width $\Gamma(l^{+}l^{-})$ with $\Gamma_{cf}$ and without $\Gamma$ correction factor in keV for $c\bar{c}$ bound system}
	\begin{ruledtabular}
	\begin{tabular}{cccccccc}
		States & Ours $\Gamma$ & Ours $\Gamma_{cf}$ & Exp\cite{exp} & \cite{70} & \cite{79} & \cite{55} & \cite{73} 
		\\
		\hline 
		$1^{3}S_{1}$ & 11.47 & 6.63 & 5.53$\pm$0.10 & 6.02 & 3.623 & 6.60 & 3.93
		\\
		$2^{3}S_{1}$ & 5.84 & 3.37 & 2.33$\pm$0.04 & 2.33 & 1.085 & 2.40 & 1.78
		\\
		$3^{3}S_{1}$ & 3.76 & 2.17 & 0.86$\pm$0.07 & 1.55 & 0.748 & 1.42 & 1.11
		\\
		$4^{3}S_{1}$ & 2.60 & 1.50 & 0.58$\pm$0.07 & 1.19 & 0.599 & 0.97 & 0.78
		\\
		$5^{3}S_{1}$ & 1.81 & 1.05 &  & 0.97 & 0.508 & 0.70 & 0.57
		\\
		$6^{3}S_{1}$ & 1.22 & 0.71 &  & 0.82 & 0.442 & 0.49 & 0.42
		\\
		\hline
		$1^{3}D_{1}$ & 0.19 & 0.11 & 0.26$\pm$0.02 & 0.14 & 0.113 & 0.031 & 0.22
		\\
		$2^{3}D_{1}$ & 0.31 & 0.18 & 0.48$\pm$0.22 & 0.22 & 0.166 & 0.037 & 0.30
		\\
		$3^{3}D_{1}$ & 0.35 & 0.20 &  & 0.26 & 0.211 & 0.044 & 0.33
		\\
		$4^{3}D_{1}$ & 0.33 & 0.19 &  & 0.20 &  &  & 0.31
		\\
		$5^{3}D_{1}$ & 0.28 & 0.16 &  & 0.23 &  &  & 0.28
	\end{tabular}
	\end{ruledtabular}
\end{table}

The di-photonic decay widths $\Gamma(\gamma \gamma)$ of charmonium states with $(\Gamma_{cf})$ and without $(\Gamma)$ the correction factor are evaluated for $\eta_{c}(nS)$, $\chi_{c0}(nP)$ and $\chi_{c2}(nP)$, and are prseneted in Table \ref{tab:14} in comparison with experimental data and other theoretical models. The di-photonic decay width of $\eta_{c}(1S)$ is higher by a factor of two and for $\eta_{c}(2S)$ it is comparable with the experimental value \cite{exp}. $X(3940)$ is considered as $\eta_{c}(3S)$ in our model and we predict its di-photonic decay width to be $3.74$ keV. The di-photonic decay widths of $\chi_{c0}(1P)$ in our model is higher compared to experimental measurement. The di-photonic decay of $\chi_{c0}(3860)$ and $X(4500)$ are not observed, but it is observed for $X(4350)$. As per our assignments we predict their values to be $3.73$ keV, $1.59$ keV and $2.19$ keV, respectively. The di-photonic decay widths of $\chi_{c2}(1P)$ in our model is in excellent agreement with the experimental value \cite{exp}. The di-photonic decay of $\chi_{c2}(3930)$ has been observed experimentally but it is not measured yet, we predict its value to be $0.51$ keV.

The tri-photonic decay widths $\Gamma(\gamma \gamma \gamma)$ of charmonium states with $(\Gamma_{cf})$ and without $(\Gamma)$ the correction factor are evaluated for $\psi(nS)$ in Table \ref{tab:15} and are compared with experimental data and other theoretical models. The tri-photonic decay width of $J/\psi(1S)$ in our model is larger when the correction term is absent and highly suppressed when it is included compared to the experimentally observed value.

\begin{table}
	\caption{\label{tab:14} Di-photonic decay width $\Gamma(\gamma \gamma)$ with $\Gamma_{cf}$ and without $\Gamma$ correction factor in keV for $c\bar{c}$ bound system}
	\begin{ruledtabular}
	\begin{tabular}{cccccccc}
		States & Ours $\Gamma$ & Ours $\Gamma_{cf}$ & Exp\cite{exp} & \cite{100} & \cite{79} & \cite{102} & \cite{55} 
		\\
		\hline 
		$1^{1}S_{0}$ & 16.41 & 12.00 & 5.15$\pm$0.35 & 7.231 & 6.621 & 7.5 & 8.5 
		\\
		$2^{1}S_{0}$ & 8.03 & 5.87 & 2.15$\pm$2.1 & 5.507 & 2.879 & 2.9 & 2.4 
		\\
		$3^{1}S_{0}$ & 5.11 & 3.74 &  & 4.971 & 2.444 & 2.5 & 0.88 
		\\
		$4^{1}S_{0}$ & 3.51 & 2.57 &  & 4.688 & 2.291 & 1.8 & 
		\\
		$5^{1}S_{0}$ & 2.44 & 1.78 &  & 4.507 & 2.213 &  & 
		\\
		$6^{1}S_{0}$ & 1.64 & 1.20 &  & 4.377 & 2.161 &  & 
		\\
		\hline
		$1^{3}P_{0}$ & 4.37 & 4.44 & 2.20$\pm$0.22 & 8.982 & 2.015 & 10.8 & 2.5 
		\\
		$2^{3}P_{0}$ & 3.67 & 3.73 &  & 9.111 & 2.349 & 6.7 & 1.7 
		\\
		$3^{3}P_{0}$ & 2.85 & 2.89 &  & 9.104 & 2.773 & 6.5 & 1.2 
		\\
		$4^{3}P_{0}$ & 2.15 & 2.19 &  & 9.076 &  &  &  
		\\
		$5^{3}P_{0}$ & 1.57 & 1.59 &  & 9.047 &  &  &  
		\\
		\hline
		$1^{3}P_{2}$ & 1.01 & 0.58 & 0.56$\pm$0.04 & 1.069 & 0.229 & 0.27 & 0.31 
		\\
		$2^{3}P_{2}$ & 0.88 & 0.51 &  & 1.084 & 0.267 & 0.39 & 0.23 
		\\
		$3^{3}P_{2}$ & 0.69 & 0.40 &  & 1.0846 & 0.315 & 0.66 & 0.17 
		\\
		$4^{3}P_{2}$ & 0.54 & 0.31 &  & 1.080 &  &  & 
		\\
		$5^{3}P_{2}$ & 0.39 & 0.23 &  & 1.077 &  &  & 
	\end{tabular}
	\end{ruledtabular}
\end{table}

\begin{table}
	\caption{\label{tab:15} Tri-photonic decay width $\Gamma(\gamma \gamma \gamma)$ with $\Gamma_{cf}$ and without $\Gamma$ correction factor in eV for $c\bar{c}$ bound system}
	\begin{ruledtabular}
	\begin{tabular}{ccccccc}
		States & Ours $\Gamma$ & Ours $\Gamma_{cf}$ & Exp\cite{exp} & \cite{72} & \cite{79} & \cite{103} 
		\\
		\hline 
		$1^{3}S_{1}$ & 6.103 & 0.031 & 1.08$\pm$0.032 & 1.022 & 3.94748 & 0.68 
		\\
		$2^{3}S_{1}$ & 3.106 & 0.016 &  & 0.900 & 1.64365 & 0.25 
		\\
		$3^{3}S_{1}$ & 2.003 & 0.010 &  & 0.857 & 1.38752 & 0.17 
		\\
		$4^{3}S_{1}$ & 1.384 & 0.007 &  & 0.832 & 1.29756 & 0.13 
		\\
		$5^{3}S_{1}$ & 0.966 & 0.005 &  & 0.815 & 1.25145 & 0.11 
		\\
		$6^{3}S_{1}$ & 0.653 & 0.003 &  & 0.801 & 1.2205 &  
	\end{tabular}
	\end{ruledtabular}
\end{table}

The di-gluonic decay widths $\Gamma(gg)$ of charmonium states with $(\Gamma_{cf})$ and without $(\Gamma)$ the correction factor are evaluated for $\eta_{c}(nS)$, $\chi_{c0}(nP)$, $\chi_{c2}(nP)$, and $\eta_{c}(nD)$ in Table \ref{tab:16} and  $\chi_{c2}(nF)$, $\chi_{c3}(nF)$, $\chi_{c4}(nF)$ and $\eta_{c}(nG)$ in Table \ref{tab:17} and are compared with experimental data and other theoretical models. The di-gluonic decay widths evaluation for $\eta_{c}(1S)$, $\eta_{c}(2S)$, $\chi_{c0}(1P)$, and $\chi_{c2}(1P)$ are in good agreement with the experimental values and show substantial improvement over other models. There is currently no experimental data on the di-gluonic decay for $\eta_{c}(nD)$, $\chi_{c2}(nF)$, $\chi_{c3}(nF)$, $\chi_{c4}(nF)$ and $\eta_{c}(nG)$.

\begin{table*}
	\caption{\label{tab:16} Di-gluonic decay width $\Gamma(gg)$ with $\Gamma_{cf}$ and without $\Gamma$ correction factor of $S$ and $P$ states in MeV and $D$ states in keV for $c\bar{c}$ bound system}
	\begin{ruledtabular}
	\begin{tabular}{cccccccc}
		States & Ours $\Gamma$ & Ours $\Gamma_{cf}$ & Exp\cite{exp} & \cite{100} & \cite{79} & \cite{104} & \cite{105} 
		\\
		\hline 
		$1^{1}S_{0}$ & 21.33 & 29.42 & 28.6$\pm$2.2 & 35.909 & 36.587 & 42.287 & 17.4$\pm$2.8 
		\\
		$2^{1}S_{0}$ & 10.43 & 14.39 & 14.7$\pm$0.7 & 27.345 & 15.910 & 27.281 & 8.3$\pm$1.3 
		\\
		$3^{1}S_{0}$ & 6.64 & 9.16 &  & 24.683 & 13.507 & 25.216 &  
		\\
		$4^{1}S_{0}$ & 4.56 & 6.29 &  & 23.281 & 12.662 & 24.415 & 
		\\
		$5^{1}S_{0}$ & 3.17 & 4.37 &  & 22.379 & 12.230 & 24.230 & 
		\\
		$6^{1}S_{0}$ & 2.14 & 2.95 &  & 23.736 & 11.941 & 24.318 & 
		\\
		\hline
		$1^{3}P_{0}$ & 5.68 & 9.94 & 10.0$\pm$0.6 & 37.919 & 9.274 & 14.19 & 14.3$\pm$3.6 
		\\
		$2^{3}P_{0}$ & 4.77 & 8.36 &  & 38.462 & 10.810 & 24.973 &  
		\\
		$3^{3}P_{0}$ & 3.71 & 6.48 &  & 38.433 & 12.758 & 33.876 &  
		\\
		$4^{3}P_{0}$ & 2.79 & 4.89 &  & 38.315 &  & 42.591 &  
		\\
		$5^{3}P_{0}$ & 2.04 & 3.57 &  & 39.191 &  &  &  
		\\
		\hline
		$1^{3}P_{2}$ & 1.31 & 1.09 & 1.97$\pm$0.11 & 3.974 & 0.945 & 2.914 & 1.71$\pm$0.21 
		\\
		$2^{3}P_{2}$ & 1.14 & 0.94 &  & 4.034 & 1.101 & 5.099 &  
		\\
		$3^{3}P_{2}$ & 0.91 & 0.74 &  & 4.028 & 1.300 & 6.867 &  
		\\
		$4^{3}P_{2}$ & 0.69 & 0.58 &  & 4.016 &  & 8.747 & 
		\\
		$5^{3}P_{2}$ & 0.52 & 0.43 &  & 4.003 &  &  & 
		\\
		\hline
		$1^{1}D_{2}$ & 7.71 &  &  &  & 12.460 &  & 110
		\\
		$2^{1}D_{2}$ & 8.94 &  &  &  & 21.679 &  & 
		\\
		$3^{1}D_{2}$ & 8.19 &  &  &  & 31.757 &  & 
		\\
		$4^{1}D_{2}$ & 6.81 &  &  &  &  &  & 
		\\
		$5^{1}D_{2}$ & 5.24 &  &  &  &  &  & 
	\end{tabular}
	\end{ruledtabular}
\end{table*}

\begin{table*}
	\caption{\label{tab:17} Di-gluonic decay width $\Gamma(gg)$ without $\Gamma$ correction factor of $F$ and $G$ states in keV for $c\bar{c}$ bound system}
	\begin{ruledtabular}
	\begin{tabular}{cccccccc}
		States & Ours $\Gamma$ & States & Ours $\Gamma$ & States & Ours $\Gamma$ & States & Ours $\Gamma$ \\
		\hline 
		$1^{3}F_{2}$ & 11.42 & $1^{3}F_{3}$ & 1.25 & $1^{3}F_{4}$ & 1.28 & $1^{1}G_{4}$ & 0.17 
		\\
		$2^{3}F_{2}$ & 16.42 & $2^{3}F_{3}$ & 1.79 & $2^{3}F_{4}$ & 1.81 & $2^{1}G_{4}$ & 0.28 
		\\
		$3^{3}F_{2}$ & 16.89 & $3^{3}F_{3}$ & 1.83 & $3^{3}F_{4}$ & 1.84 & $3^{1}G_{4}$ & 0.31 
		\\
		$4^{3}F_{2}$ & 15.03 & $4^{3}F_{3}$ & 1.63 & $4^{3}F_{4}$ & 1.63 & $4^{1}G_{4}$ & 0.29 
		\\
		$5^{3}F_{2}$ & 11.99 & $5^{3}F_{3}$ & 1.30 & $5^{3}F_{4}$ & 1.30 & $5^{1}G_{4}$ & 0.24 
	\end{tabular}
	\end{ruledtabular}
\end{table*}

\begin{table*}
	\caption{\label{tab:18} Tri-gluonic decay width $\Gamma(ggg)$ with $\Gamma_{cf}$ and without $\Gamma$ correction factor of $S$, $P$ and $D$ states in keV for $c\bar{c}$ bound system}
	\begin{ruledtabular}
	\begin{tabular}{ccccccc}
		States & Ours $\Gamma$ & Ours $\Gamma_{cf}$ & Exp\cite{exp} & \cite{79} & \cite{105} & \cite{103} 
		\\
		\hline 
		$1^{3}S_{1}$ & 252.69 & 178.86 & 59.35$\pm$2.01 & 269.059 & 52.8$\pm$5 & 101 
		\\
		$2^{3}S_{1}$ & 128.62 & 91.04 & 31.16$\pm$5.61 & 112.031 & 23$\pm$2.6 & 36.6 
		\\
		$3^{3}S_{1}$ & 82.95 & 58.72 &  & 94.5727 &  & 24.7 
		\\
		$4^{3}S_{1}$ & 57.29 & 40.55 &  & 88.4413 &  & 19.8 
		\\
		$5^{3}S_{1}$ & 39.99 & 28.30 &  & 85.2984 &  & 16.7 
		\\
		$6^{3}S_{1}$ & 27.06 & 19.15 &  & 83.1888 &  &  
		\\
		\hline
		$1^{1}P_{1}$ & 207.26 &  &  & 285.127 & 720$\pm$320 &  
		\\
		$2^{1}P_{1}$ & 242.90 &  &  & 420.078 &  &  
		\\
		$3^{1}P_{1}$ & 228.00 &  &  & 558.78 &  &  
		\\
		$4^{1}P_{1}$ & 197.96 &  &  &  &  &  
		\\
		$5^{1}P_{1}$ & 162.15 &  &  &  &  &  
		\\
		\hline
		$1^{3}D_{1}$ & 84.19 &  &  & 189.367 & 216 &  
		\\
		$2^{3}D_{1}$ & 110.40 &  &  & 359.346 &  &  
		\\
		$3^{3}D_{1}$ & 109.96 &  &  & 556.588 &  &  
		\\
		$4^{3}D_{1}$ & 97.85 &  &  &  &  &  
		\\
		$5^{3}D_{1}$ & 80.13 &  &  &  &  &  
		\\
		\hline
		$1^{3}D_{2}$ & 9.76 &  &  & 53.8761 & 36 &  
		\\
		$2^{3}D_{2}$ & 12.81 &  &  & 102.236 &  &  
		\\
		$3^{3}D_{2}$ & 12.79 &  &  & 158.353 &  &  
		\\
		$4^{3}D_{2}$ & 11.41 &  &  &  &  &  
		\\
		$5^{3}D_{2}$ & 0.94 &  &  &  &  &  
		\\
		\hline
		$1^{3}D_{3}$ & 39.12 &  &  & 89.7001 & 102 &  
		\\
		$2^{3}D_{3}$ & 50.91 &  &  & 170.217 &  &  
		\\
		$3^{3}D_{3}$ & 50.72 &  &  & 263.647 &  &  
		\\
		$4^{3}D_{3}$ & 45.29 &  &  &  &  &  
		\\
		$5^{3}D_{3}$ & 37.26 &  &  &  &  &  
	\end{tabular}
	\end{ruledtabular}
\end{table*}

The tri-gluonic decay widths $\Gamma(ggg)$ of charmonium states  with $(\Gamma_{cf})$ and without $(\Gamma)$ the correction factor are evaluated for $\psi(nS)$, $h_{c}(nP)$, $\psi_{1}(nD)$, $\psi_{2}(nD)$ and $\psi_{3}(nD)$ in Table \ref{tab:18} and are compared with experimental data and other theoretical models. The tri-gluonic decay widths for $J/\psi(1S)$ and $\psi(2S)$ are higher in our model than the experimentally measured values.

The photo-gluon decay widths $\Gamma(\gamma gg)$ for $\psi(nS)$ and quark-gluon decay width $\Gamma(q\bar{q} + g)$ for $\chi_{c1}(nP)$ of charmonium states with $(\Gamma_{cf})$ and without $(\Gamma)$ the correction factor are evaluated in Table \ref{tab:19} and are compared with experimental data and other theoretical models. The photo-gluon decay widths in our model for $J/\psi(1S)$ and $\psi(2S)$ are in accordance with the experimentally determined values. The quark-gluon decay width for $\chi_{c1}(nP)$ has no experimental data.

\begin{table*}
	\caption{\label{tab:19} Photo-gluon decay width $\Gamma(\gamma gg)$ of $S$ states and quark-gluon decay width $\Gamma(q\bar{q} + g)$ of $P$ states in keV for $c\bar{c}$ bound system}
	\begin{ruledtabular}
	\begin{tabular}{cccccc}
		States & Ours $\Gamma$ & Ours $\Gamma_{cf}$ & Exp\cite{exp} & \cite{79} & \cite{103} 
		\\
		\hline 
		$1^{3}S_{1}$ & 23.79 & 11.20 & 8.15$\pm$1.17 & 8.996 & 6.18 
		\\
		$2^{3}S_{1}$ & 12.11 & 5.70 & 3.03$\pm$0.93 & 3.746 & 2.25 
		\\
		$3^{3}S_{1}$ & 7.81 & 3.68 &  & 3.162 & 1.52 
		\\
		$4^{3}S_{1}$ & 5.39 & 2.54 &  & 2.957 & 1.22 
		\\
		$5^{3}S_{1}$ & 3.76 & 1.77 &  & 2.852 & 1.03 
		\\
		$6^{3}S_{1}$ & 2.55 & 1.19 &  & 2.782 &  
		\\
		\hline 
		$1^{3}P_{1}$ & 249.56 &  &  & 342.152 &  
		\\
		$2^{3}P_{1}$ & 292.34 &  &  & 504.093 &  
		\\
		$3^{3}P_{1}$ & 274.36 &  &  & 670.536 &  
		\\
		$4^{3}P_{1}$ & 238.15 &  &  &  &  
		\\
		$5^{3}P_{1}$ & 194.98 &  &  &  &  
	\end{tabular}
	\end{ruledtabular}
\end{table*}

\begin{table*}
	\caption{\label{tab:20} $S-D$ mixed states of $c\bar{c}$ with masses of mixed states $M_{\phi}$ and $M_{\phi'}$ in MeV and their di-leptonic decay width $\Gamma_{\phi}$ and $\Gamma_{\phi'}$ in keV}
	\begin{ruledtabular}
	\begin{tabular}{ccccccc}
		$S-D$ & $M_{S}$ & $\theta$ & $M_{\phi}$ & $M_{exp}$ & $\Gamma_{\phi}$ & $\Gamma_{exp}$ 
		\\
		States & $M_{D}$ &  & $M_{\phi'}$ & \cite{exp} & $\Gamma_{\phi'}$ & \cite{exp}
		\\
		\hline 
		$2S$ & 3690.0 & -6.41 & 3687.9 & 3686.1 & 3.17 & 2.33$\pm$0.04 \\
		$1D$ & 3817.1 &  & 3819.2 & 3773.7 & 0.29 & 0.26$\pm$0.02 
		\\
		\hline
		$3S$ & 4048.2 & 45.13 & 4029.1 & 4039.6 & 0.99 & 0.86$\pm$0.07 \\
		$2D$ & 4123.3 &  & 4142.8 & 4191.0 & 0.56 & 0.48$\pm$0.22 
		\\
		\hline
		$4S$ & 4294.4 & -24.15 & 4254.3 & 4222.1 & 1.10 &  
		\\
		$3D$ & 4346.0 &  & 4386.2 & 4374.0 & 0.83 &  
		\\
		\hline
		$5S$ & 4468.2 & -16.11 & 4458.8 & 4415.0 & 0.59 & 0.58$\pm$0.07 \\
		$4D$ & 4510.8 &  & 4520.3 & 4484.0 & 0.50 &  
	\end{tabular}
	\end{ruledtabular}
\end{table*}

\begin{table*}
	\caption{\label{tab:21} Our assignments of $c\bar{c}$ states with masses $M_{exp}$ and $M_{cal}$ in MeV and di-leptonic decay widths $\Gamma_{exp}^{ee}$ and $\Gamma_{cal}^{ee}$ in keV}
	\begin{ruledtabular}
	\begin{tabular}{cccccc}
		States & Assignment & $M_{exp}$\cite{exp} & $M_{cal}$ & $\Gamma_{exp}^{ee}$\cite{exp} & $\Gamma_{cal}^{ee}$ \\
		\hline 
		$\eta_{c}(1S)$ & $\eta_{c}(1S)$ & 2983.9$\pm$0.4 & 2986.3 &  &  
		\\
		$J/\psi$ & $\psi(1S)$ & 3096.9$\pm$0.006 & 3094.1 & 5.53$\pm$0.10 & 6.63 
		\\
		$\chi_{c0}(1P)$ & $\chi_{c0}(1P)$ & 3414.71$\pm$0.3 & 3434.4 &  &  
		\\
		$\chi_{c1}(1P)$ & $\chi_{c1}(1P)$ & 3510.67$\pm$0.05 & 3514.2 &  &  
		\\
		$h_{c}(1P)$ & $h_{c}(1P)$ & 3525.37$\pm$0.14 & 3517.2 &  &  
		\\
		$\chi_{c2}(1P)$ & $\chi_{c2}(1P)$ & 3556.17$\pm$0.07 & 3556.2 &  &  
		\\
		$\eta_{c}(2S)$ & $\eta_{c}(2S)$ & 3637.7$\pm$1.1 & 3633.1 &  &  
		\\
		$\psi(2S)$ & $\psi(2S)-\psi(1D)$ & 3686.1$\pm$0.06 & 3687.9 & 2.33$\pm$0.04 & 3.17 
		\\
		$\psi(3770)$ & $\psi(2S)-\psi(1D)$ & 3778.7$\pm$0.7 & 3819.2 & 0.26$\pm$0.02 & 0.29 
		\\
		$\psi_{2}(3823)$ & $\psi_{2}(1D)$ & 3823.5$\pm$0.5 & 3830.3 &  & 
		\\
		$\psi(3842)$ & $\psi_{3}(1D)$ & 3842.71$\pm$0.16$\pm$0.12 & 3829.3 &  & 
		\\
		$\chi_{c0}(3860)$ & $\chi_{c0}(2P)$ & $3862_{-32-13}^{+26+40}$ & 3857.5 &  &  
		\\
		$\chi_{c1}(3872)$ & $\chi_{c1}(2P)$ & 3871.65$\pm$0.06 & 3924.7 &  &  
		\\
		$\chi_{c2}(3930)$ & $\chi_{c2}(2P)$ & 3922.5$\pm$1.0 & 3965.2 &  &  
		\\
		$X(3940)$ & $\eta_{c}(3S)$ & $3862_{-6}^{+7}$$\pm$6 & 4011.9 &  &  
		\\
		$\psi(4040)$ & $\psi(3S)-\psi(2D)$ & 4039.6$\pm$4.3 & 4029.1 & 0.86$\pm$0.07 & 0.99
		\\
		$X(4140)$ & $\chi_{c1}(3P)$ & 4146.5$\pm$3.0 & 4206.9 &  &  
		\\
		$\psi(4160)$ & $\psi(3S)-\psi(2D)$ & 4191$\pm$5 & 4142.8 & 0.48$\pm$0.22 & 0.56
		\\
		$\psi(4230)$ & $\psi(4S)-\psi(3D)$ & 4222.1$\pm$2.3 & 4254.3 &  & 1.10
		\\
		$X(4274)$ & $\chi_{c1}(3P)$ & $4286_{-6}^{+7}$ & 4206.9 &  &  
		\\
		$X(4350)$ & $\chi_{c0}(4P)$ & $4350.6_{-5.1}^{+4.6}$$\pm$0.7 & 4366.6 &  &  
		\\
		$\psi(4360)$ & $\psi(4S)-\psi(3D)$ & 4374$\pm$7 & 4386.2 &  & 0.83
		\\
		$\psi(4415)$ & $\psi(5S)-\psi(4D)$ & 4415$\pm$5 & 4458.8 & 0.58$\pm$0.07 & 0.59
		\\
		$X(4500)$ & $\chi_{c0}(5P)$ & 4474$\pm$3$\pm$3 & 4525.1 &  &  
		\\
		$Y(4500)$ & $\psi(5S)-\psi(4D)$ & 4484$\pm$13.3$\pm$24.1 & 4520.3 &  & 0.50
		\\
		$Y(4660)$ & $\psi_{1}(5D)$ & 4630$\pm$6 & 4631.5 &  & 0.16
	\end{tabular}
	\end{ruledtabular}
\end{table*}

The masses and di-leptonic decay widths of $S-D$ mixed states are presented in Table \ref{tab:20} for $2S-1D$, $3S-2D$, $4S-3D$, and $5S-4D$ charmonium states. The mixed states are also assigned to experimentally observed states. The $\psi(2S)$ and $\psi(3770)$ predominantly represent charmonium states with a small $2S-1D$ mixing component. This proposition also has been put forth in ref. \cite{70,95}. The $\psi(4040)$ and $\psi(4160)$ do not align perfectly as pure charmonium resonances due to their mass and decay width values. Therefore, $\psi(4040)$ and $\psi(4160)$ are suggested to be $3S-2D$ mixed states with substantial mixing. This interpretation is consistent with predictions in ref. \cite{70}. The branching fraction for di-leptonic decays of $\psi(4040)$ and $\psi(4160)$ yield $B[\Gamma(\psi(4040) \rightarrow e^{+}e^{-})] = (1.07 \pm 0.16) \times 10^{-5}$ and $B[\Gamma(\psi(4160) \rightarrow e^{+}e^{-})] = (6.9 \pm 3.3) \times 10^{-6}$, respectively. Utilizing these values, the total decay width of $\psi(4040)$ and $\psi(4160)$ is calculated to be $93 \pm 14$ MeV and $81 \pm 39$ MeV, respectively. In a unified Fano-like interference model, it has been suggested that $Y(4008)$, $Y(4260)$ and $Y(4360)$ may not be true resonances \cite{106}, while $Y(4330)$ and $Y(4390)$ can be reduced to a single resonance $\psi(4230)$ \cite{107}. The $\psi(4230)$ is considered to be $4S-3D$ mixed states with a substantial mixing component, in agreement with the findings in ref. \cite{107}. $B[\Gamma(\psi(4230) \rightarrow \mu^{+}\mu^{-})]$ is measured to be $(3.2 \pm 2.9) \times 10^{-5}$ \cite{exp}. Considering the $B[\Gamma(\psi(4230) \rightarrow e^{+}e^{-})] \approx B[\Gamma(\psi(4230) \rightarrow \mu^{+}\mu^{-})]$, the total decay width of $\psi(4230)$ is calculated to be $34 \pm 31$ MeV. In our model, the $\psi(4360)$ is proposed to be the partner of $\psi(4230)$ with a mass of $4386.2$ MeV. Instead of $\psi(4360)$, a new particle $\psi(4380)$ is suggested as the partner of $\psi(4230)$, with a central mass value of $4384$ MeV and di-leptonic decay width of $0.257$ keV \cite{107}. The di-leptonic decay width in our model for $\psi(4360)$ is estimated to be $0.83$ keV. Experimental verification of $\psi(4380)$, clarification on whether $\psi(4380)$ and $\psi(4360)$ are identical states, and determination of the di-leptonic decay width for $\psi(4360)$ and $\psi(4380)$ are necessary to resolve the uncertainties regarding the partner of $\psi(4230)$. The $\psi(4415)$ and $Y(4500)$ are suggested to be $5S-4D$ mixed states with a notable mixing component \cite{78,107}. Our interpretation of $5S-4D$ mixed states also aligning with ref. \cite{78,107}. The di-leptonic decay width of $\psi(4415)$ as mixed state is evaluated to be $0.59$ keV, consistent with experimental value \cite{exp}. The di-leptonic decay width of $Y(4500)$ is predicted to be $0.50$ keV. From the branching fraction $B[\Gamma(\psi(4415) \rightarrow e^{+}e^{-}) = (9.4 \pm 3.2) \times 10^{-6}$ \cite{exp}, we obtain the total decay width of $\psi(4415)$ to be $63 \pm 21$ MeV. Upon considering the $S-D$ mixing, our final assignments are presented in Table \ref{tab:21}.

\clearpage

\section{\label{sec:Conclusion} Conclusion}

This comprehensive investigation utilizes a screened potential model for the computation of the mass spectrum and decay widths of the $c\bar{c}$ bound system within a relativistic framework. The results of our study can offer valuable insights into the structure of various charmonium-like $XYZ$ states, providing a comparison with experimental data and other theoretical models. Our model provides predictions for $S,P,D,F$ and $G$ wave spectra as well as decay constants, $E1$ transitions, $M1$ transitions, annihilation decay widths, as well as the masses and di-leptonic decay widths of $S-D$ mixed states. In interpreting the observed charmonium-like $XYZ$ states, this analysis provides insight into their potential classification as either conventional charmonium states or exotic structures. Through rigorous comparison with experimental data, each state assignment is carefully scrutinized. Highlighting the model's ability to discern between different interpretations is proposed in the literature. Our study suggests that several charmonium states such as $\psi(3770)$, $\psi(4040)$, $\psi(4160)$, $\psi(4230)$, $\psi(4360)$, $\psi(4415)$, and $Y(4500)$, exhibit significant $S-D$ mixing, impacting their mass and decay properties. This observation contributes to understanding the complexity of charmonium structures and provides a reference for further exploration of these states.

\nocite{*}

\bibliography{CharmoniumArxiveReference}

\end{document}